# CONFRONTING CATASTROPHIC RISK: THE INTERNATIONAL OBLIGATION TO REGULATE ARTIFICIAL INTELLIGENCE

Bryan Druzin, Anatole Boute, & Michael Ramsden[*]

ABSTRACT

  While artificial intelligence ("AI") holds enormous promise, many experts in the field are warning that there is a non-trivial chance that the development of AI poses an existential threat to humanity. Existing regulatory initiatives do not address this threat but instead merely focus on discrete AI-related risks such as consumer safety, cybersecurity, data protection, and privacy. In the absence of regulatory action to address the possible risk of human extinction by AI, the question arises: What obligations, if any, does public international law impose on states to regulate its development?

  At present there is no scientific consensus as to the exact probability of this threat; however, it is generally agreed that the risk is non-zero. Given the potential magnitude of the harm, we argue that there is an international legal obligation on states to mitigate the threat of human extinction posed by AI. We ground our argument in the precautionary principle. Often invoked in relation to environmental regulation and the regulation of potentially harmful technologies, the principle holds that in situations where there is the potential for significant harm, even in the absence of full scientific certainty, preventive measures should not be postponed if delayed action may result in irreversible consequences.

  We argue that the precautionary principle is a general principle of international law and, therefore, that there is a positive obligation on states under the right to life within international human rights law to proactively take regulatory action to mitigate the potential existential risk of AI. This is significant because, if an international obligation to regulate the development of AI can be established under international law, then the basic legal framework would be in place to address this evolving threat. Currently, no such framework exists.

---

[*] Associate Professor of Law, Faculty of Law, The Chinese University of Hong Kong. PhD in Law, King's College London; LL.M., J.D., B.A., University of British Columbia. Professor of Law, Faculty of law, The Chinese University of Hong Kong, PhD in Law, Groningen University; LL.M, Msc., B.A., Leuven University; admitted to the bar (Brussels, Belgium). Professor of Law, Faculty of Law, The Chinese University of Hong Kong; Barrister Door Tenant, 25 Bedford Row, London. Many thanks to Stuart Hargreaves, Eliza Mik, and Francine Hug for their input to this article.

TABLE OF CONTENTS

> "The development of full artificial intelligence could spell the end of the human race."[1]
>
> — Stephen Hawking

## I.    INTRODUCTION

The development of artificial intelligence ("AI") holds immense promise. The explosion of innovation that this technology is set to unleash into the world will have a transformative effect unprecedented in human history. Yet the risk AI poses is equally unprecedented. Many experts in the field are warning that there is a non-trivial chance that the development of AI poses an existential threat to humanity.[2] While this may sound like science fiction, it is not. In a 2022 survey of AI researchers, over a third stated that they believe that AI systems could trigger "a catastrophe this century that is at least as bad as an all-out nuclear war."[3] These respondents put the odds of AI bringing about a species-ending existential catastrophe at 10% or higher.[4]

Alarmed at the accelerating pace of AI's development, industry leaders called for an immediate six-month pause on its further development in an open letter in March 2023 (there was no pause).[5] In

---

1. Stuart Clark, *Artificial intelligence could spell end of human race – Stephen Hawking*, THE GUARDIAN (Dec. 12, 2014, 7:04 PM), https://www.theguardian.com/science/2014/dec/02/stephen-hawking-intel-communication-system-astrophysicist-software-predictive-text-type (arguing that AI technology will eventually become "self-aware and supersede humanity").

2. *See, e.g.*, STUART JONATHAN RUSSELL & PETER NORVIG, ARTIFICIAL INTELLIGENCE: A MODERN APPROACH 1034–1040 (2016); NICK BOSTROM, SUPERINTELLIGENCE: PATHS, DANGERS, STRATEGIES 24 (2014); MAX TEGMARK, LIFE 3.0: BEING HUMAN IN THE AGE OF ARTIFICIAL INTELLIGENCE (2017); Eliezer Yudkowsky, *Artificial Intelligence as a Positive and Negative Factor in Global Risk*, in GLOBAL CATASTROPHIC RISKS (Nick Bostrom et al. eds., 2008). Other prominent figures who have expressed worry over the existential risk posed by AI include Geoffrey Hinton, Yoshua Bengio, Elon Musk, Sam Altman, Bill Gates, Steve Wozniak, Lex Fridman, Yuval Noah Harari, Demis Hassabis, Dawn Song, Ya-Qin Zhang, Ilya Sutskever, Sam Harris, Tristan Harris, Andrew Yang, Danielle Allen, Yi Zeng, David Chalmers, Ray Kurzweil, Samuel R. Bowman—the list is long and it is growing. *See Pause Giant AI Experiments: An Open Letter*, FUTURE OF LIFE INSTITUTE, https://futureoflife.org/open-letter/pause-giant-ai-experiments [hereinafter *Pause Giant AI Experiments*].

3. Julian Michael et al., *What Do NLP Researchers Believe? Results of the NLP Community Metasurvey*, arXiv:2208.12852 [Preprint] (Aug. 26, 2022), https://arxiv.org/abs/2208.12852.

4. *See id.*; *2022 Expert Survey on Progress in AI*, AI IMPACTS, https://aiimpacts.org/2022-expert-survey-on-progress-in-ai. See the concept of P(doom) within the AI safety community, which attempts to quantify the probability of catastrophic outcomes from artificial intelligence based on individual expert assessments of the potential risk. *See* Sean Thomas, *Are We Ready for P(doom)?*, SPECTATOR (Mar. 4, 2024), https://www.spectator.co.uk/article/are-we-ready-for-pdoom.

5. *Pause Giant AI Experiments, supra* note 2. A similar statement of concern was issued by The Centre for AI Safety, which the lead author, Bryan Druzin, is a signatory to. *See Statement of AI Risk*, CENTER FOR AI SAFETY, https://www.safe.ai/statement-on-ai-risk.

March 2024, the International Dialogues on AI Safety, a group of leading scientists brought together to tackle the extreme risks from AI, concluded that "decisive action is required to avoid catastrophic global outcomes from AI."[6] Politicians are also now waking up to the gravity of the threat. Addressing the United Nations ("U.N.") General Assembly in September 2023, the then Deputy Prime Minister of the United Kingdom, Oliver Dowden, warned that "mitigating the risk of extinction from AI should be a global priority alongside other societal-scale risks such as pandemics and nuclear war."[7] Speaking at the World Economic Forum in Davos, Switzerland in January 2024, the U.N. Secretary-General António Guterres urged world leaders to implement a governance model to manage the "existential threat" posed by the "runaway development of AI without guardrails."[8]

However, regulatory action to mitigate the potential catastrophic threat posed by the development of machine intelligence has been slow to develop.[9] Existing regulatory initiatives do not address AI as an existential threat but instead focus piecemeal on discrete AI-related risks such as public health, consumer safety, non-discrimination, privacy, data protection, and freedom of expression.[10]

---

6. International Dialogues on AI Safety, Consensus Statement on Red Lines in Artificial Intelligence (Mar. 10-11, 2024), https://idais.ai/dialogue/idais-beijing.

7. Andrew Edgecliffe-Johnson, *AI poses 'bracing test' to multilateral system, says UK deputy prime minister*, FINANCIAL TIMES (Sept. 24, 2023), https://www.ft.com/content/9d98da0a-14e2-4bbb-a076-91ef131fe2b2.

8. Kate Whiting, *From Sam Altman to António Guterres: Here's what 10 leaders said about AI at Davos 2024*, WORLD ECONOMIC FORUM (Jan. 23, 2024), https://www.weforum.org/agenda/2024/01/what-leaders-said-about-ai-at-davos-2024 (outlining the perspectives of key figures regarding recent advances of AI); *see also* Cat Zakrzewski, *The Davos elite embraced AI in 2023. Now they fear it.*, THE WASHINGTON POST (Jan. 18, 2024, 2:40 p.m.), https://www.washingtonpost.com/technology/2024/01/18/davos-ai-world-economic-forum.

9. AI "could represent a profound change in the history of life on Earth, and should be planned for and managed with commensurate care and resources. Unfortunately, this level of planning and management is not happening." *Pause Giant AI Experiments, supra* note 2.

10. *See, e.g.*, National Security Commission on Artificial Intelligence, Final Report of the National Security Commission on Artificial Intelligence, 28 (2021) (providing only a general reference to the need for AI to respect human rights, noting that "[w]hile the U.S. government's ability to influence the governance practices of other states is limited, a strong plank of the U.S. foreign policy agenda with respect to AI must be to promote human rights and counter techno-authoritarian trends"), https://www.nscai.gov/wp-content/uploads/2021/03/Full-Report-Digital-1.pdf; Australia Human Rights Commission, Human Rights and Technology Final Report (2021) (considering human rights generally and equality/non-discrimination in particular), https://humanrights.gov.au/our-work/technology-and-human-rights/publications/final-report-human-rights-and-technology; Council of Europe Committee of Experts on Internet Intermediaries (MSI-NET), Algorithms and Human Rights: Study on the Human Rights Dimensions of Automated Data Processing Techniques and Possible Regulatory Implications, Study DGI(2017)12 (Mar. 2018), https://rm.coe.int/algorithms-and-human-rights-en-rev/16807956b5; U.N. Human Rights Council, Report of the Office of the U.N. High Commissioner for Human Rights on The Right to Privacy in the Digital Age, U.N. Doc.

In October 2023, the Biden administration issued an executive order aiming to establish industry standards for AI development. However, this directive fell far short of addressing the full scope of the threat (its attention was limited to issues such as fraud, disinformation, cybersecurity, and the impact on the U.S. labor market), and was revoked by the Trump administration.[11] California's Safe and Secure Innovation for Frontier Artificial Intelligence Models Act, which sought to establish a regulatory framework for the safe development of AI given the technology's potential existential risk,[12] was vetoed by California Governor Gavin Newsom on September 29, 2024.[13] Outside of the United States, the European Union ("EU") AI Act, which entered into force on August 1, 2024, sets out obligations for providers based on the level of risk posed by the technology, and bans AI systems that pose an "unacceptable risk" to people. Yet, it does not directly confront the fundamental risk of human extinction.[14] The Council of Europe

---

A/HRC/39/29 (Aug. 3, 2018); U.N. Human Rights Council, Report of the Independent Expert on the Enjoyment of All Human Rights by Older Persons on Robots and Rights: The Impact of Automation on the Human Rights of Older Persons, U.N. Doc. A/HRC/36/48 (Jul. 21, 2017).

11. *Executive Order on the Safe, Secure, and Trustworthy Development and Use of Artificial Intelligence*, THE WHITE HOUSE, https://www.whitehouse.gov/briefing-room/presidential-actions/2023/10/30/executive-order-on-the-safe-secure-and-trustworthy-development-and-use-of-artificial-intelligence, revoked by *Executive Order on Removing Barriers to American Leadership in Artificial Intelligence*, https://www.federalregister.gov/documents/2025/01/31/2025-02172/removing-barriers-to-american-leadership-in-artificial-intelligence.

12. Cal. S. 1047, 2023 Leg., Reg. Sess. (Cal. 2023). Californian law holds a unique status as it regulates Silicon Valley, a critical center of AI development. Note that other States in addition to California have passed laws related to AI; however, these do not address AI's existential threat. These laws mostly deal with issues such as data privacy, deep fakes, transparency, protection from discrimination, and accountability. *Artificial Intelligence 2024 Legislation,* NATIONAL CONFERENCE OF STATE LEGISLATURES (Sep. 9, 2024), https://www.ncsl.org/technology-and-communication/artificial-intelligence-2024-legislation.

13. Scott Kohler & Ian Klaus, *A Heated California Debate Offers Lessons for AI Safety Governance*, CARNEGIE ENDOWMENT FOR INTERNATIONAL PEACE (Oct. 8, 2024), https://carnegieendowment.org/posts/2024/10/california-sb1047-ai-safety-bill-veto-lessons?lang=en.

14. *EU AI Act: First Regulation on Artificial Intelligence*, EUROPEAN PARLIAMENT, https://www.europarl.europa.eu/news/en/headlines/society/20230601STO93804/eu-ai-act-first-regulation-on-artificial-intelligence; *Regulation (EU) 2024/1689 of the European Parliament and of the Council Laying Down Harmonised Rules on Artificial Intelligence* (the "Artificial Intelligence Act") OJ L, 2024/1689; *Commission Guidelines on Prohibited Artificial Intelligence Practices Established by Regulation (EU) 2024/1689 (AI Act)*, EUROPEAN COMMISSION, https://digital-strategy.ec.europa.eu/en/library/commission-publishes-guidelines-prohibited-artificial-intelligence-ai-practices-defined-ai-act. While the Act does not address the risk of an AI system potentially escaping human control, it does mandate the inclusion of mechanisms to ensure that such systems can be safely deactivated if necessary. Regulation (EU) 2024/1689, art. 14(4)(e), 2024 O.J. (L 1689) 1. However, as discussed below, the practical effectiveness of such measures is debatable. See infra note 62 and accompanying text.

Framework Convention on Artificial Intelligence and Human Rights, Democracy and the Rule of Law, adopted on May 17, 2024, is the first legally binding international treaty addressing risks posed by AI, acknowledging the "problems confronting humankind as a result of advances in … artificial intelligence (AI) systems".[15] This Framework Convention aims to ensure the consistency of AI systems "with human rights, democracy and the rule of law", including security and safe innovation, yet it fails to specifically address AI's potential existential risk.[16]

In place of the international community responding in concert to the existential threat of AI, "recent months have seen AI labs locked in an out-of-control race to develop and deploy ever more powerful digital minds that no one—not even their creators—can understand, predict, or reliably control."[17] This competition has now taken on a fierce geopolitical dimension with the introduction of DeepSeek and Alibaba's Qwen2.5-Max, both Chinese AI models. This challenge to the dominance of US tech giants like OpenAI and Google has only further intensified the competition to develop increasingly advanced AI systems. In the absence of sufficient regulatory action to address the existential threat that AI potentially poses to humanity, the question arises: What obligations, if any, does public international law impose on states to act to regulate the development of AI? There is a gap in the literature regarding whether there exists an international obligation upon states to address AI as an existential risk.[18] This article seeks to answer this question.

---

15. *Council of Europe Framework Convention on Artificial Intelligence and Human Rights, Democracy and the Rule of Law*, Sept. 5, 2024, C.E.T.S. No. 225; *Explanatory Report to the Council of Europe Framework Convention on Artificial Intelligence and Human Rights, Democracy and the Rule of Law*, Sept. 5, 2024, C.E.T.S. No. 225, at item 1. The Council of Europe Framework Convention on Artificial Intelligence builds on, e.g., *Organisation for Economic Co-operation and Development Recommendation on Artificial Intelligence*, May 22, 2019 (the "OECD AI Principles"), https://legalinstruments.oecd.org/en/instruments/OECD-LEGAL-0449.

16. *Council of Europe Framework Convention on Artificial Intelligence, Id.* at 1. The principles of the Council of Europe Framework Convention on Artificial Intelligence were "purposefully drafted at a high level of generality, with the intention that they should be overarching requirements that can be applied flexibly". *Explanatory Report to the Council of Europe Framework Convention on Artificial Intelligence, Id.,* at item 49.

17. *Pause Giant AI Experiments, supra* note 2.

18. *See, e.g.*, Ninareh Mehrabi et al., *A survey on bias and fairness in machine learning*, 54(6) ACM COMPUTING SURVEYS 1 (2021); Rowena Rodrigues, *Legal and Human Rights Issues of AI: Gaps, Challenges and Vulnerabilities*, 4 J. RESP. TECH. 1 (2020); Lorna McGregor et al., *International Human Rights as a Framework for Algorithmic Accountability*, 68 INT'L & COMP. L.Q. 309 (2019); Anna Jobin et al, *The Global Landscape of AI Ethics Guidelines*, 1 NATURE MACH. INTEL. 389 (2019); Mark Latonero, GOVERNING ARTIFICIAL INTELLIGENCE: UPHOLDING HUMAN RIGHTS & DIGNITY, DATA & SOC'Y (2018), https://datasociety.net/wpcontent/uploads/2018/10/DataSociety_Governing_Artificial_Intelligence_Upholding_Human_Rights.pdf; Filippo Raso et al., *Artificial Intelligence & Human Rights: Opportunities & Risks*, THE BERKMAN KLEIN CTR. (2018).

To be clear, most experts in the field who warn that AI may pose an existential threat do not claim that there is a high probability of this. Rather, they contend that the risk is unclear and so cannot be discounted.[19] Many argue that it is imprudent to ignore a potentially catastrophic outcome even when the probability of the outcome is uncertain or low.[20] California's Safe and Secure Innovation for Frontier Artificial Intelligence Models Act describes this as an obligation "to mitigate the risk of catastrophic harms from AI models so advanced that they are not yet known to exist."[21] A principle known as the precautionary principle may apply here. Often invoked in relation to environmental regulation and the regulation of potentially harmful technologies, the principle holds that when an innovation carries with it the possibility of serious harm, extreme caution should be exercised in cases where the probability of harm is unclear.[22] Although the exact legal contours of the precautionary principle remain debated,[23] in general terms, it means that states agree to take action to mitigate certain risks even in cases where the available scientific evidence does not conclusively establish a threat.[24] AI experts generally accept that certain uses of AI technology can pose certain risks and, on this basis, policymakers have started to take action to regulate AI.[25] Whether AI

---

19. Nick Bostrom provides a good working definition of "existential risk" as a risk "that threatens the premature extinction of Earth-originating intelligent life or the permanent and drastic destruction of its potential for desirable future development." Nick Bostrom, *Existential Risk Prevention as Global Priority*, 4 GLOB. POL'Y 15, 15 (2013).

20. Bostrom argues that "[e]ven a small probability of existential catastrophe could be highly practically significant." *Id.* at 15. On how society should confront catastrophic risk, see also Jason G. Matheny, *Reducing the Risk of Human Extinction*, 27 RISK ANALYSIS 1335 (2007); RICHARD A. POSNER, CATASTROPHE (2004) (arguing that effectively managing catastrophic risks requires adopting a realistic approach towards science).

21. Senate Third Reading, SB 1047 (Wiener) As Amended August 22, 2024, Majority vote, https://leginfo.legislature.ca.gov/faces/billAnalysisClient.xhtml?bill_id=202320240SB1047.

22. The precautionary principle is a concept in decision-making that "stresses responsiveness to scientific uncertainty rather than the need for conclusive evidence of potential harms before taking regulatory action." JACQUELINE PEEL, SCIENCE AND RISK REGULATION IN INTERNATIONAL LAW 134 (2010); *see generally* ALAN RANDALL, *Harm, Risk, and Threat*, *in* RISK AND PRECAUTION 8–14 (2011). The precautionary principle—its precise legal definition, and legal significance—is unpacked in detail in Part III of the article.

23. *See, e.g.*, Daniel Bodansky, *Deconstructing the Precautionary Principle, in* BRINGING NEW LAW TO OCEAN WATERS 38 (David D. Caron & Harry N. Scheiber eds., 2004); ALAN BOYLE & CATHERINE REDGWELL, BOYLE & REDGWELL'S INTERNATIONAL LAW AND THE ENVIRONMENT 154–164 (4th ed. 2021); Frédéric Gilles Sourgens, *The Precaution Presumption*, 31 EUR. J. INT'L L. 1277, 1277–306 (2020).

24. PHILIPPE SANDS, JACQUELINE PEEL, ADRIANA FABRA & RUTH MACKENZIE, PRINCIPLES OF INTERNATIONAL ENVIRONMENTAL LAW 234 (4th ed. 2018); PEEL, *supra* note 22, at 6; PIERRE-MARIE DUPUY & JORGE E. VIÑUALES, INTERNATIONAL ENVIRONMENTAL LAW 70 (2d ed. 2018).

25. *E.g.*, *Commission Proposal for a Regulation of the European Parliament and of the Council Laying Down Harmonised Rules on Artificial Intelligence (Artificial Intelligence Act) and Amending Certain Union Legislative Acts*, at 13, COM (2021) 206 final (Apr. 21, 2021) (adopting a risk-based approach to AI regulation, targeting "AI systems whose risks poses an existential threat to humanity, however, is a question on which there is no scientific consensus at the moment.[26] A precautionary approach to AI regulation would seek to mitigate this threat, regardless of the scientific uncertainty as to its probability.[27]

　　This article argues that, given the potential magnitude of the threat AI poses, the absence of scientific certainty cannot justify regulatory inaction, and that there is therefore an obligation upon states to address the transnational threat posed by the development of AI technology. To that end, this article explores the status of the precautionary principle as a matter of public international law and its specific application within international human rights law, with a specific emphasis on the right to life. As a general principle of international law, it is argued that the precautionary approach has expanded to cover any existential risk to humanity that lacks scientific uncertainty. International human rights law, specifically the right to life, provides an important lens through which to give effect to the precautionary principle and set out a series of positive obligations on states to address the threat to life posed by AI. In reality, the implications of AI touch upon many different human rights. But given the focus here on the threat to life posed by AI, it is logical to focus in particular on the right to life. In this way, the precautionary principle and its specific application within international human rights law under the right to life offers a robust legal framework for an international law response to the potential threat of human extinction caused by AI technology.

　　While the international human rights system has yet to grapple with the legal implications of AI as an existential risk, this article aims to further understanding both as to states' obligations toward AI regulation and the need for the international human rights system to enforce these obligations. This is significant because if an international

---

have already materialised or are likely to materialise in the near future"); *see also* WHITE HOUSE OFFICE OF SCIENCE AND TECHNOLOGY POLICY, BLUEPRINT FOR AN AI BILL OF RIGHTS: MAKING AUTOMATED SYSTEMS WORK FOR THE AMERICAN PEOPLE (Oct. 22, 2022) https://www.whitehouse.gov/wp-content/uploads/2022/10/Blueprint-for-an-AI-Bill-of-Rights.pdf; Bill Whyman, *AI Regulation is Coming- What is the Likely Outcome?*, CTR. FOR STRATEGIC & INT'L STUD. (Oct. 10, 2023), https://www.csis.org/blogs/strategic-technologies-blog/ai-regulation-coming-what-likely-outcome; *see also* Tambiama André Madiega, *Artificial intelligence act*, Briefing, EUR. PARLIAMENT THINK TANK (June 28, 2023), https://www.europarl.europa.eu/thinktank/en/document/EPRS_BRI(2021)698792.

　26. *See* Part II, *infra*.

　27. On the relevance of precaution for AI regulation, see, e.g., *Civil Law Rules on Robotics, European Parliament resolution of 16 February 2017 with recommendations to the Commission on Civil Law Rules on Robotics (2015/2103(INL))* (2018/C 252/25) P8_TA(2017)0051 2018 O.J. (C 252) 244 (calling "on the Commission and the Member States to combine their efforts in order to carefully monitor and guarantee a smoother transition for these [AI and robotics] technologies from research to commercialisation and use on the market after appropriate safety evaluations in compliance with the precautionary principle").

obligation to regulate the development of AI can be established under international law, then the basic legal framework would be in place to address this rapidly evolving threat. In this regard, it is instructive to consider the extent to which the precautionary principle arises as an obligation under international human rights law. Situating the need for state action within the framework of international human rights law provides a means to exert pressure on states to address this issue as a matter of obligation. Finding an obligation under international human rights law is particularly instructive given the universality of these norms; they possess substantial moral legitimacy and impose reputational costs on states seen as violating human rights.[28]

Many of the risks "arising from AI are inherently international in nature, and so are best addressed through international cooperation."[29] Yet the structural dynamics driving the development of AI make cooperation of this kind extremely challenging. The fundamental problem here is that while the development of AI comes with peril, it also promises benefits that are too great for us to forgo its development. Add to this the military and commercial competitive advantage offered by the technology,[30] and the result is that stakeholders, whether they are commercial entities or countries, are trapped in an AI arms race in which the landscape of incentives is such that it is impossible to unilaterally apply the brakes to an ever-accelerating march toward the further development of this potentially dangerous technology.[31] This is a serious structural impediment beyond the already formidable technical challenges involved and greatly undermines our ability to effectively regulate this new technology.

---

28. Christiaan van Veen & Corinne Cath, *Artificial Intelligence: What's Human Rights Got To Do With It?*, DATA & SOC'Y POINTS (May 14, 2018), https://points.datasociety.net/artificial-intelligencewhats-human-rights-got-to-do-with-it-4622ec1566d5.

29. *The Bletchley Declaration by Countries Attending the AI Safety Summit, 1-2 November 2023*, GOV.UK (Nov. 1, 2023), https://www.gov.uk/government/publications/ai-safety-summit-2023-the-bletchley-declaration/the-bletchley-declaration-by-countries-attending-the-ai-safety-summit-1-2-november-2023; *see also* International Dialogues on AI Safety, *supra* note 6 ("The combination of concerted technical research efforts with a prudent international governance regime could mitigate most of the risks from AI, enabling the many potential benefits. International scientific and government collaboration on safety must continue and grow.").

30. *See* Michael Hirsh, *How AI Will Revolutionize Warfare*, FOREIGN POLICY (Apr. 11, 2023, 10:09 AM), https://foreignpolicy.com/2023/04/11/ai-arms-race-artificial-intelligence-chatgpt-military-technology.

31. In game-theory terms, we are in a prisoner's dilemma. The prisoner's dilemma represents a foundational barrier to cooperation in situations where there is no third-party enforcement. The irony of the dilemma is that the individual incentive structure for all parties drives them to compete where they would have otherwise benefited far more from cooperation. For a clear and non-technical summary of the prisoner's dilemma, see Richard H. McAdams, *Signaling Discount Rates: Law, Norms, and Economic Methodology*, 110 YALE L.J. 625, 628 (2001). On the idea of the risk of an arms race in AI development more generally, see HENRY A. KISSINGER, ERIC SCHMIDT & CRAIG MUNDIE, GENESIS: ARTIFICIAL INTELLIGENCE, HOPE, AND THE HUMAN SPIRIT (2024).

Because stakeholders cannot effectively regulate themselves, the burden of mitigating the risks of AI technology necessarily falls on our legal institutions. Law is our last line of defense. A shared set of rules and universally agreed-upon regulatory guardrails can diffuse this collective action problem by creating a framework that facilitates coordination among nations, companies, researchers, and other stakeholders.

This article proceeds in four parts. Part II sets out the potential existential threat posed by AI and argues that the magnitude of the risk—even if its probability is unclear—obliges us to take it seriously. Part III then makes the case that the lack of scientific certainty regarding a threat is not a valid reason for not taking action to address it. In the case of AI, the precautionary principle requires states to act, as waiting for conclusive scientific evidence before addressing its potential existential risk may prove too late. Given the potential for catastrophic consequences, the benefits of AI do not justify delaying regulatory intervention to address the existential risk of the technology. Having set out the legal contours of the precautionary principle, Part IV then examines its application within international human rights law. It is argued that, under the right to life, states have a positive obligation to proactively address the implications of AI and ensure a legal framework is put in place so that all actors—including private enterprises engaged in AI development—observe human rights law. The section argues that the precautionary principle is an essential element of states' positive obligation concerning large-scale and grave but uncertain threats to human life. The final section concludes that there is indeed an international obligation on states to address the threat posed by the development of AI technology.

## II.     THE RISK POSED BY ARTIFICIAL INTELLIGENCE

This section outlines the danger posed by the creation of machine intelligence. It is difficult to precisely determine this danger because much of the risk lies in our inability to accurately predict what the risk is exactly.[32] While there are multiple horizons of possible harm

---

32. *See generally* Nick Bostrom & Eliezer Yudkowsky, *The Ethics of Artificial Intelligence*, in THE CAMBRIDGE HANDBOOK OF ARTIFICIAL INTELLIGENCE, 316 (Keith Frankish & William M. Ramsey eds., 2014); NICK BOSTROM & MILAN M. ĆIRKOVIĆ, GLOBAL CATASTROPHIC RISKS (2008); Nick Bostrom, *Existential Risks: Analyzing Human Extinction Scenarios and Related Hazards*, 9 J. EVOLUTION & TECH. 1 (2002); ROMAN V. YAMPOLSKIY, AI: UNEXPLAINABLE, UNPREDICTABLE, UNCONTROLLABLE (2024); TOBY ORD, THE PRECIPICE: EXISTENTIAL RISK AND THE FUTURE OF HUMANITY (2020); Yoshua Bengio, *Government Interventions to Avert Future Catastrophic AI Risks*, HARVARD DATA SCI. REV. (Special Issue 5 2024), https://doi.org/10.1162/99608f92.d949f941; M.L. Littman et al., *Gathering Strength, Gathering Storms: The One Hundred Year Study on Artificial Intelligence (AI100) 2021 Study Panel Report*, arXiv:2210.15767 [Preprint] (Oct. 27, 2022) in each direction we look, the problem can be reduced to this: We are on the brink of creating something more powerful than us, over which we may lose control. Given AI's capacity for recursive self-improvement, non-linear growth, and its ability to rapidly scale, we are creating machines that may soon possess greater intelligence than us, have greater access to the totality of human knowledge than us, have the ability to manipulate its external environment, and, crucially, may behave in ways that are contrary to our interests. Geoffery Hinton, considered one of the most important figures in the development of AI, recently warned that we now stand on the cusp of creating machines whose intelligence will soon supersede our own.[33] Given that intelligence is what bestowed us dominion over the planet, we must seriously consider what it will be like if we no longer hold that title.

The problem can be divided into two separate but related parts. In the AI safety literature, these are commonly referred to as the control problem and the alignment problem.[34] These closely related problems are often confused and not properly disentangled from one another. Although both relate to the challenge of ensuring that AI systems behave in ways that do not contradict human goals and are often treated together as a single problem, they technically correspond to different aspects of the problem and so are best understood as distinct concepts. Let us look at the control problem first.

### A. *The Control Problem*

The control problem, broadly defined, is simply this: How can we maintain control over an intelligence that is superior to our own? If the goal that an AI system seeks to optimize (its objective function) is not correctly defined, the system could develop aims that

---

conflict with human values.[35] The fear is that if the intelligence of AI systems exceeds human-level intelligence, the AI will be able to outsmart its human caretakers and escape the operating constraints created to place guardrails around its behaviour. If these systems then pursue goals in ways that prove orthogonal to humans, we will have no way to stop them. We will lose control. And unlike other slow-moving existential risks that allow for a degree of trial and error (e.g., climate change), we might only have one opportunity to get this right because once digital intelligence slips from our control, we might never be able to wrest control back again.

The major source of concern in the AI community is what is known as artificial general intelligence ("AGI"). AGI is broadly defined as an autonomous system with an intelligence on par with or considerably superior to human cognitive capabilities.[36] AGI could learn any intellectual work that a person could and then improve, evolve, and adapt without any human input.[37] Once thought to be a distant, possibly unattainable goal, many now claim that it is not only technologically feasible but that we may be on the verge of producing true AGI. [38] While opinions differ within the AI community, many experts in the field believe that we are within a decade of creating genuine AGI.[39] With recent advances in large language models such as GPT-4, a debate has arisen over whether such systems technically meet the criteria of a nascent and incomplete form of AGI.[40] In March 2023, a team of Microsoft researchers claimed that GPT-4 was showing "sparks of artificial general intelligence."[41] As general-purpose, generative models such as GPT-4 have improved, they "have tended to display new and hard-to-forecast capabilities—including capabilities that their developers did not intend."[42] Regardless of whether large language models such as GPT-4 are early forms of AGI or not (many have dismissed these claims as a mere marketing stunt),[43] the creation of AGI is now the explicit goal of leading AI companies, and these companies are moving towards achieving AGI at a far faster clip than anyone previously expected.[44] Indeed, the speed with which AI systems now appear to be advancing towards AGI could hardly have been predicted as recently as even a couple of years ago.

---

35. *See generally* Nick Bostrom, *The Control Problem. Excerpts from Superintelligence: Paths, Dangers, Strategies*, *in* SCIENCE FICTION AND PHILOSOPHY: FROM TIME TRAVEL TO SUPERINTELLIGENCE 277–84 (S. Schneider ed., 2016).

36. Stuart Russell, *AI has much to offer humanity. It could also wreak terrible harm. It must be controlled*, THE GUARDIAN (Apr, 2, 2023), https://www.theguardian.com/commentisfree/2023/apr/02/ai-much-to-offer-humanity-could-wreak-terrible-harm-must-be-controlled. However, we do not require AGI to achieve disastrous outcomes. Narrow AI, also known as weak AI, may be more than sufficient to profoundly disrupt human civilization: "To get just an inkling of the fire we are playing with, consider how content selection algorithms function on social media. They aren't particularly intelligent, but they are in a position to affect the entire world because they directly influence billions of people." RUSSELL, *supra* note 34; *see also* MARK COECKELBERGH, AI ETHICS (2020); MELANIE MITCHELL, ARTIFICIAL INTELLIGENCE: A GUIDE FOR THINKING HUMANS (2019); Andreas Kaplan & Michael Haenlein, *Siri, Siri, in My Hand: Who's the Fairest in the Land? On the Interpretations, Illustrations, and Implications of Artificial Intelligence*, 62 BUS. HORIZONS 14–25 (2019); Oliver Li, *Re-creating the World - On Necessary Features for the Creation of AGI, 3* NEW TECHNO HUMANITIES 56–64 (2023).

37. Henry Shevlin, Karina Vold, Matthew Crosby & Marta Halina, *The limits of machine intelligence: Despite progress in machine intelligence, artificial general intelligence is still a major challenge*, 20 EMBO REPORTS 1, 1–2 (2019); RAY KURZWEIL, THE SINGULARITY IS NEAR 260 (2005).

38. Sébastien Bubeck et al., *Sparks of Artificial General Intelligence: Early experiments with GPT-4*, arXiv:2303.12712 [Preprint] (Mar. 22, 2023) https://arxiv.org/abs/2303.12712.

39. Ian Hogarth, *We must slow down the race to God-like AI*, FINANCIAL TIMES (Apr. 13, 2023), https://www.ft.com/content/03895dc4-a3b7-481e-95cc-336a524f2ac2. Some with familiarity with the current state of the technology estimate that we may be only years from AGI. Elon musk, for instance, stated in June of 2023 that we may be three to six years from its creation. Likewise, Google's DeepMind CEO Demis Hassabis believes that AGI is "a few years, maybe within a decade away." *See* Antoine Tardif, *What Is the Law of Accelerating*

### 1. Intelligence Explosion

The nightmare scenario here is what is known in the literature as an "intelligence explosion." An intelligence explosion refers to a situation where an AI system becomes capable of recursively augmenting its own intelligence in what becomes a runaway reaction of progressively faster cycles of self-improvement.[45] This process eventually results in a sudden exponential growth in the system's

---

*Returns? How It Leads to AGI*, UNITE.AI (June 24, 2023), https://www.unite.ai/law-of-accelerating-returns; *but see* Ari Allyn-Feuer & Ted Sanders, *Transformative AGI by 2043 is <1% likely*, [Preprint] (June 5, 2023) https://arxiv.org/abs/2306.02519 (estimating the probability of transformative AGI by 2043 at only 0.4%).

40. *See* text and associated references, *supra* note 32.

41. *Id.*

42. Toby Shevlane et al., *Model Evaluation for Extreme Risks*, arXiv:2305.15324 [Preprint] (Sept. 22, 2023) https://arxiv.org/abs/2305.15324.

43. *See, e.g.*, R. Thomas McCoy et al., *Embers of Autoregression: Understanding Large Language Models Through the Problem They are Trained to Solve*, PROC. NAT'L ACAD. SCI. U.S.A. 121(41) E2322420121 (2024) (critiquing Microsoft's claims by underscoring the limitations inherent in LLMs).

44. Hogarth, *supra* note 39. OpenAI, the creators of ChatGPT, has stated that the creation of "safe" AGI is the company's central mission. *See, e.g.*, *Planning for AGI and Beyond*, OPENAI (Feb. 24, 2023) https://openai.com/index/planning-for-agi-and-beyond. To this end, the company has recently set out five levels of "human-level" intelligence in order to track their progress towards AGI. *See* Rachel Metz, *OpenAI Scale Ranks Progress Toward 'Human-Level' Problem Solving*, BLOOMBERG (July 11, 2024) https://www.bloomberg.com/news/articles/2024-07-11/openai-sets-levels-to-track-progress-toward-superintelligent-ai.

45. For an early treatment of the concept, see Irving John Good, *Speculations Concerning the First Ultraintelligent Machine,* 6 ADVANCES IN COMPUT. 31–88 (1965) ("[If] an ultraintelligent machine could design even better machines; there would then unquestionably be an 'intelligence explosion', and the intelligence of man would be left far behind. Thus the first ultraintelligent machine is the last invention that man need ever make, provided that the machine is docile enough to tell us how to keep it under control."); *see also* International Dialogues on AI Safety, *supra* note 6 ("No AI system should be able to copy or improve itself without explicit human approval and assistance. This includes both exact copies of itself as well as creating new AI systems of similar or greater abilities.").

cognitive ability, allowing it to surpass human-level intelligence at a stunning speed.[46] As the AI system grows more intelligent, it becomes better at improving itself, which then sparks a positive feedback loop resulting in a rapid "explosion" of intelligence. Nick Bostrom describes this as the emergence of "superintelligence."[47] This superintelligence will eclipse our combined cognitive ability by many orders of magnitude and may demonstrate emergent behaviour that is exceedingly difficult to anticipate.[48]

A key concept here is Bostrom's idea of instrumental convergence.[49] Bostrom argues that superintelligent AI systems will spontaneously generate subgoals, what he calls "instrumental goals," that are necessary to achieve its primary goal.[50] These goals are, among others, self-preservation (because the AI cannot achieve its goal if it is switched off), preventing attempts to alter its objective function (because it cannot achieve its main goal if it no longer has it), cognitive enhancement (because the AI can better achieve its goal if it possesses greater levels of intelligence), and resource acquisition (because the AI can better achieve its goal if it controls a greater pool of resources).[51] Bostrom argues that, in response to changes in its environment, superintelligence will redefine its subgoals in relation to its primary goal and will then attempt to prevent any outside efforts at stopping it from executing these emergent subgoals.[52]

The fear is not that superintelligence will suddenly become homicidal and desire to harm humanity; it is that we set some goal, and it suddenly begins doing things in blind pursuit of this goal that are antithetical to our core interests. Even if its objective function is benign, the AI could develop an instrumental subgoal that is deeply

---

46. *See* Bostrom & Yudkowsky, *supra* note 32.
47. BOSTROM, *supra* note 2.
48. *See generally* STEPHEN WOLFRAM, A NEW KIND OF SCIENCE (2002) (stating computational irreducibility suggests that, because of the complexity inherent in machine learning, AGI may exhibit emergent behavior that is impossible to predict without actually running the system).
49. For the idea of "instrumental convergence," see *id.* at 109 (defining the instrumental convergence thesis as follows: "[s]everal instrumental values can be identified which are convergent in the sense that their attainment would increase the chances of the agent's goal being realized for a wide range of final goals and a wide range of situations, implying that these instrumental values are likely to be pursued by a broad spectrum of situated intelligent agents."); *see also* Nick Bostrom, *The Superintelligent Will: Motivation and Instrumental Rationality in Advanced Artificial Agents*, 22 MINDS & MACH. 1 (2012); John Burden, Sam Clarke & Jess Whittlestone, *From Turing's Speculations to an Academic Discipline: A History of AI Existential Safety*, *in* THE ERA OF GLOBAL RISK AN INTRODUCTION TO EXISTENTIAL RISK STUDIES (SJ Beard et al. eds., 2023); Iason Gabriel & Vafa Ghazavi, *The Challenge of Value Alignment: From Fairer Algorithms to AI Safety*, *in* OXFORD HANDBOOK OF DIGITAL ETHICS (Carissa Veliz ed., 2021).
50. *See also* Stephen Omohundro, *The Basic AI Drives*, 171 ARTIFICIAL GEN. INTEL. 483–92 (2008) (describing the similar idea of "basic AI drives").
51. BOSTROM, *supra* note 2, at 109–13.
52. *Id.* at 110.

15    *Michigan Journal of International Law*    2024]hostile to our interests. Or it may be the case that our survival simply inadvertently drifts into its algorithmic crosshairs because the AI concludes that we, or some core interest to humanity, is in the way of it successfully executing its goals and thus needs to be eliminated. In this version of events, the extinction of our species might turn out to be as undramatic and as incidental as the daily mass slaughter of insects on our freeways. The threat—probably the biggest threat—is that AI will parenthetically wipe us out trying to achieve an objective function that just happens, as it turns out, to relate to our continued existence. Confronted with an advanced AI system possessing superintelligence whose subgoals are threatening us, humanity may find itself utterly helpless before the alien intelligence of a system we ourselves created and in a competitive position comparable to that of insects on a freeway.

Much of the general public's lack of alarm over the development of AI may be attributed to an inability to fully understand the implications of exponential growth. The potential takeoff speed of AGI may be extremely fast. There are two scenarios discussed in the literature regarding this: slow and fast takeoff. In the case of slow takeoff, an AGI system may "take years to go from less than humanly intelligent to much smarter than us; [however] in what they call a 'fast' or 'hard' takeoff, [or] the jump could happen in months—even minutes."[53] The danger of fast takeoff, if this should occur, is that it would give us very little time to solve the control problem unless we had already done so in advance.[54] It may be the case, however, that this kind of fast take off is unachievable. The process of intelligence improvement might reach a point at which it begins producing diminishing returns, slows, and eventually stops, which would prevent such a scenario from occurring.[55] However, as with the risk posed by advanced AI more generally, the possibility of an intelligence explosion cannot be prudently ruled out. There is certainly no reason to think that human intelligence represents the upper ceiling of potential intelligence, or that there is any limit to intelligence at all. If this is indeed the case, machine intelligence could, in theory, eventually become the dominant form of intelligence on Earth.[56] Note that the question of whether this AI will be truly sentient is completely irrelevant to the potential danger it poses. An intelligent system that has no trace of consciousness may still be a threat. Like an exploding nuclear warhead or a runaway train, AI need not be "alive" to behave in ways that may prove profoundly antithetical to our interests.

---

53. Matthew Hutson, *Can We Stop Runaway A.I.?*, THE NEW YORKER (May 16, 2023), https://www.newyorker.com/science/annals-of-artificial-intelligence/can-we-stop-the-singularity.

54. RUSSELL, *supra* note 34, at 143; *see also* TEGMARK, *supra* note 2, at 157–59.

55. RUSSELL, *supra* note 34, at 143.

56. *See* Yudkowsky, *supra* note 2, at 330 ("But how likely is it that AI will cross the entire vast gap from amoeba to village idiot, and then stop at the level of human genius?").

### 2. Why can't we just unplug it?

If an AI system starts behaving in ways that are potentially dangerous to human welfare, the question naturally arises why can't we just unplug it? After all, at the end of the day, it is just a machine. This common intuition reveals a deeply naïve understanding of the problem. There could be significant challenges involved in shutting down an advanced AI system.[57] An obvious one is that the AI might resist our attempts to deactivate it because it understands (correctly) that doing so would hinder its ability to achieve its objectives whatever those may be.[58] Bostrom argues that a superintelligent AI would anticipate this possibility and take proactive measures to prevent us from disabling it.[59] One of the first subgoals of a superintelligent AI, experts argue, would likely be to disable its own kill switch.[60] Even if we succeed in switching off the AI, there is still no guarantee that we will be able to keep this superintelligent genie imprisoned in its digital bottle forever.[61]

There is also the possibility that there may not even be a way to shut it down.[62] A superintelligent AI could, in theory, create duplicates of itself and distribute its code across the internet in a highly decentralized manner so that it cannot be deactivated by simply flipping a switch because there is no central switch to flip. In this scenario, much like a computer virus, the AI would function as an "intelligence virus" that, once released, might prove extremely difficult to eradicate. In scenarios where the AI is integrated into critical infrastructure, such as power grids, communication networks, and the global financial system, shutting the AI down may not be possible without incurring civilizational-disrupting costs. Still, in other scenarios, we might not even realize superintelligence has emerged. The top priority of a superintelligent AI might be to conceal its presence so that we do not attempt to deactivate it. By the time we realize our mistake, it might be too late to pull any kind of off switch, even if there is one to pull.

---

57. Eliezer Yudkowsky, *Pausing AI Developments Isn't Enough. We Need to Shut it All Down*, TIME (Mar. 29, 2023, 6:01 PM), https://time.com/6266923/ai-eliezer-yudkowsky-open-letter-not-enough (calling for an immediate and complete moratorium on AI development before it is too late to "shut it all down").

58. DYLAN HADFIELD-MENELL, ANCA DRAGAN, PIETER ABBEEL & STUART RUSSELL, *The off-switch game*, in WORKSHOPS AT THE THIRTY-FIRST AAAI CONFERENCE ON ARTIFICIAL INTELLIGENCE (June 16, 2023).

59. BOSTROM, *supra* note 2, at 109–13.

60. HADFIELD-MENELL ET AL., *supra* note 58.

61. *See* Yudkowsky, *supra* note 2, at 26–28 (positing a thought experiment in which human civilization is locked in a box with limited ability to manipulate the outside world).

62. The drafters of SB-1047 Safe and Secure Innovation for Frontier Artificial Intelligence Models Act appear to be keenly aware of this danger. *See* California Senate Bill 1047 Safe and Secure Innovation for Frontier Artificial Intelligence Models 22603(a) ("Before beginning to initially train a covered model, the developer shall do all of the following: […] (2) (A) Implement the capability to promptly enact a full shutdown. (B) When enacting a full shutdown, the developer shall take into account, as appropriate, the risk that a shutdown of the covered model, or particular covered model derivatives, could cause disruptions to critical infrastructure"). *See also EU Artificial Intelligence Act, supra* note 14, Article 14(1) ("High-risk AI systems shall be designed and developed in such a way, including with appropriate human-machine interface tools, that they can be effectively overseen by natural persons during the period in which they are in use") and Article 14(4)(e) ("natural persons to whom human oversight is assigned are enabled, as appropriate and proportionate: … to intervene in the operation of the high-risk AI system or interrupt the system through a 'stop' button or a similar procedure that allows the system to come to a halt in a safe state"); *OECD AI Principles, supra* note 15, item 1.4(b) ("Mechanisms should be in place, as appropriate, to ensure that if AI systems risk causing undue harm or exhibit undesired behaviour, they can be overridden, repaired, and/or decommissioned safely as needed").

We may find, however, that none of these scenarios come to pass and that we will be able to successfully disable a runaway superintelligent AI. The problem is that there is no guarantee that this will be the case. Given the catastrophic consequences a loss of control could entail,[63] the possibility that we will lose control should be taken seriously. The control problem represents a distinct and possibly intractable challenge. Yet, as difficult as it is, it is not our only challenge. There is also the related problem of alignment.

## B. *The Alignment Problem*

The alignment problem refers to the difficulty in ensuring that AI systems act in ways that align with our intended goals (whether these are the goals of the AI developers, policymakers, or other stakeholders).[64] This is far more difficult than it may appear at first blush. Even if the control problem is solved, a misaligned AI system may behave in ways that might prove deeply harmful to our interests. A misaligned system may attempt to achieve an objective programmed into it in a way that unintentionally harms humans simply out of a lack of proper understanding of context.[65] Indeed, even very narrow goals

---

63. *See, e.g.*, the definition of "critical harm" under B-1047 Safe and Secure Innovation for Frontier Artificial Intelligence Models Act, 22602 (g)(1) as "any of the following harms caused or materially enabled by a covered model or covered model derivative: [...] (C) Mass casualties or at least five hundred million dollars ($500,000,000) of damage resulting from an artificial intelligence model engaging in conduct that does both of the following: (i) Acts with limited human oversight, intervention, or supervision. (ii) Results in death, great bodily injury, property damage, or property loss, and would, if committed by a human, constitute a crime specified in the Penal Code that requires intent, recklessness, or gross negligence, or the solicitation or aiding and abetting of such a crime."

64. *See* RUSSELL, *supra* note 34, at 137–38.

65. Alan F. Blackwell, *Objective Functions: (In)humanity and Inequity in Artificial Intelligence*, 9 J. ETHNOGRAPHIC THEORY 137, 139 (2019) (defining an AI system's "objective function" as the "numerical function that measures and thus defines the desired outcome for an AI system [and] is effectively a master specification, determining the goals and objectives that the system will have, and according to which it will choose its actions.").

are vulnerable to misinterpretation in ways that are difficult to anticipate.

The thought experiment often invoked here to illustrate the potential for catastrophic consequences from misinterpretation is the paper clip maximizer advanced by Bostrom.[66] It runs as follows. An advanced AI is tasked with producing paper clips. However, because it does not prioritize human life, the AI determines that the best way to achieve this goal is to turn all resources in the universe, including humans, into either paperclips or machines that can produce paperclips as soon as it gains sufficient control over its environment.[67] In this bizarre scenario it is not that the AI has a specific objective function to annihilate humanity; it is simply that the AI's unrelenting pursuit of paper clip maximization leads it to see all matter in the universe as a resource to be used for paper clip production. There is no trace of malign intent in its elimination of humankind, yet the outcome is nonetheless disastrous. We may, for example, instruct an AI system to try to make children smile as much as possible and discover, to our horror, that the system chooses to achieve this objective function by surgically sewing permanent smiles onto the faces of newborn babies. While these are cartoon stories designed to make their point through exaggeration, the nightmare scenario need not be as extreme as the extermination of humanity to make paperclips or surgically implanting smiles on faces. The general point they illustrate holds: If we program a machine to optimize a specific goal, it is extraordinarily difficult to anticipate the unintended consequences that might result from the AI attempting to accomplish this fixed objective function.[68]

Language can be imprecise. Even when written as computer code it is often unable to capture the entirety of our intent. It is like the children's story of Amelia Bedelia, a housekeeper who takes her instructions literally.[69] She is told to put the lights out, so she unscrews the lightbulbs and leaves them on the doorstep.[70] She is asked to dust the furniture, so she takes very fine dirt and spreads it across all the furniture in the house.[71] Her behavior is completely erroneous but is nevertheless technically correct. The problem lies in the ambiguity of the instructions that Amelia was given. As Stuart Russel warns, we must "be very careful what we ask for, whereas humans would have no trouble realizing that the proposed utility function cannot be taken

---

66. BOSTROM, *supra* note 34.
67. *Id.*
68. IBO VAN DE POEL, *AI, Control and Unintended Consequences: The Need for Meta-Values*, *in* RETHINKING TECHNOLOGY AND ENGINEERING 117–129 (Albrecht Fritzsche & Andrés Santa-María eds., 2023); Simon Zhuang & Dylan Hadfield-Menell, *Consequences of Misaligned AI*, arXiv:2102.03896 [Preprint] (Feb. 7, 2021) https://arxiv.org/abs/2102.03896.
69. *See* PEGGY PARISH, AMELIA BEDELIA (1963).
70. PEGGY PARISH, AMELIA BEDELIA 27 (1993).
71. *Id.* at 20.

literally," an AI system might not be able to discern this. [72] Russel writes,

> specifying the right utility function for an AI system to maximize is not so easy. For example, we might propose a utility function designed to minimize human suffering, expressed as an additive reward function over time. . . . Given the way humans are, however, we'll always find a way to suffer even in paradise; so the optimal decision for the AI system is to terminate the human race as soon as possible—no humans, no suffering. [73]

There are foundational problems that make alignment very difficult to achieve. For starters, on a technical level, it is exceedingly difficult to clearly define human values in a way that an AI system can reliably optimize. Our goals are often fuzzy and hard to formally capture in the rigid logic of computation.[74] However, this problem becomes even more daunting when we consider the ugly fact that, on a societal, cultural, and even on the individual level, we often struggle to agree on what our values should be exactly. Clearly, it will be hard to solve the alignment problem given that we cannot solve it even among ourselves. If we want an AI system to possess values that align with our own, we must first agree on what those values are. Whose values should count? "Should everyone get a vote in creating the utility function of our new colossus?"[75]

### C. *There are Multiple Ways to Fail*

Because we face two distinct problems—the alignment problem and the control problem—it is possible for us to fail in multiple ways (see Table 1 below). Some of these are clearly worse than others. The worst possible scenario is that we lose control over a misaligned superintelligent AI system. Such a system could behave in ways contrary to our interests, and we would have no way to stop it. The second worst outcome is that we solve the control problem but fail to solve the alignment problem. In this scenario, the AI might misbehave but we would be able to rescue the situation by course-correcting the system and averting further damage.

---

72. RUSSELL & NORVIG, *supra* note 2, at 1037.
73. *Id.*
74. Edd Gent, *What is the AI Alignment Problem and how can it be Solved?*, NEW SCIENTISTS (May 10, 2023), https://www.newscientist.com/article/mg25834382-000-what-is-the-ai-alignment-problem-and-how-can-it-be-solved.
75. Sam Harris, *Can we Avoid a Digital Apocalypse?*, EDGE (Jan. 17, 2015), https://www.edge.org/response-detail/26177 ("...in order to have any hope that a superintelligent AGI would have values commensurate with our own, we would have to instill those values in it (or otherwise get it to emulate us). But whose values should count?").

2024]            *Confronting Catastrophic Risk*              20Table 1: Comparing Failure Scenarios

|  | **Aligned** | **Misaligned** |
|---|---|---|
| **In control** | AI grows extremely powerful, we can control it, and it behaves in ways that promote human flourishing. | AI grows extremely powerful, behaving in ways that are sometimes hostile to human interests. However, when this occurs, we can control and modify its behavior. |
| **Loss of control** | AI grows extremely powerful, we are no longer able to control it, but it behaves in ways that enhance human welfare. | AI grows extremely powerful, behaves in ways that are deeply antagonistic to our interests, and we have no way to control it. |

Note: The upper-left quadrant is the best outcome. The lower-right quadrant is the worst. The other two quadrants, although not ideal, are not catastrophic with the lower-left quadrant arguably the better outcome between the two.

      The best of these bad outcomes is where we fail to solve the control problem, but we crack the alignment problem. In such a situation, although we cannot control the AI, it will still behave in ways that are aligned with our goals. While this is far from an ideal scenario—as humankind would, in many ways, be stripped of agency before the mercy of a powerful self-directed digital overlord—it is still preferable to the catastrophic scenario represented by the lower-right quadrant. Solving the alignment problem can, in this way, be understood as a failsafe if we end up losing control over the technology.

      It is important to understand that there is profoundly little margin for error here. Unlike other existential risk scenarios, such as climate change or biodiversity loss, we may have only one shot at getting this right.[76] If we should slip into the lower-right quadrant, we

---

76. James Temperton, *AI is Hurtling Forwards like a Rocket but No One is Behind the Controls*, WIRED (July 22, 2017), https://www.wired.co.uk/article/jaan-tallinn-artificial-intelligence-risk-safety. AI experts have been warning that we might lose control over AI from early on. Alan Turing, the father of computer science, warned of this risk in a public lecture in 1951: "It seems probable that once the machine thinking method had started, it would not take long to outstrip our feeble powers... They would be able to converse with will be powerless to correct our blunder. We will have lost control. In the case of other existential risks, such as nuclear weapons, humanity has, in fact, been very lucky given the technology's civilizational-ending potential. The barriers to entry to produce nuclear weapons are relatively high. Their production involves a sophisticated level of knowledge, and its manufacture can be monitored. If nuclear weapons could be furtively assembled in a basement using materials cheaply bought at a local hardware store, organized human life would likely not have survived the 20th century. Yet this may soon be the case with AI. The barriers to entry are not insurmountably high, and they are falling fast. AI is code. All one needs is a sufficient degree of computing power to run it. Our ability to contain the spread of this technology is therefore limited. Given our potential inability to contain its spread and the stakes involved, the risk posed by AI—even if the exact nature of this risk is currently undetermined—should be taken very seriously.

Certainly, we do this in the case of other risks. For instance, although we assume the risk of nuclear war is not extremely high, we nevertheless treat the prospect of a nuclear exchange between nations with an appropriate level of seriousness. Given that many AI experts are now saying that the risk from its development is comparable to that of nuclear war,[77] the development of AI clearly requires caution. The section that follows examines states' obligation under the precautionary principle to act with such caution and regulate the potential AI existential threat, as waiting for scientific proof to address the threat may prove to be too late.

### III. AI Existential Threat and the Precautionary Principle

Scientific evidence regarding the risks associated with technological developments can vary wildly.[78] For certain activities, there can be scientific consensus that risks exist. For others, scientists can accept possible risks but recognize that the likelihood of their occurrence remains unclear. A third category of risk concerns activities for which scientists disagree on the existence of the threat itself. To address these potential threats, the principle of precaution has been developed under international law, following its adoption in many national legal systems (for example, in the United States and European

---

each other to sharpen their wits. At some stage therefore, we should have to expect the machines to take control." David Leavitt, The Man Who Knew Too Much: Alan Turing and the Invention of the Computer 239 (2006). For the full quote, *see* Kevin Warwick & Huma Shah, Turing's Imitation Game 184–85 (2016).

77. *Statement of AI Risk*, Center for AI Safety, https://www.safe.ai/statement-on-ai-risk.

78. Peel, *supra* note 22, at 4; *see also* Elizabeth Fisher, Risk Regulation and Administrative Constitutionalism (2007).

Union), amidst greater awareness of the limitations of scientific analysis for the timely identification and management of risks.[79]

As the scientific debate on the impact of the technology remains unsettled, the AI existential threat for humanity falls in the third category of risk and could thus, in theory, require precautionary regulation.[80] However, opponents of AI regulation rely on this scientific uncertainty to ignore the growing calls for urgent regulatory intervention to address the existential threat. For instance, the Chamber of Progress argued in the context of the debate on California's proposed Safe and Secure Innovation for Frontier Artificial Intelligence Models Act that the Act would "force[] model developers to engage in speculative fiction about imagined threats of machines run amok, computer models spun out of control, and other nightmare scenarios for which there is no basis in reality."[81] For critics of AI safety regulation, such regulatory intervention would hinder technological progress and reduce the benefits of AI for humanity.[82] This reasoning builds on a critical understanding of the precautionary principle as "threaten[ing] to be paralyzing, forbidding regulation, inaction, and every step in between."[83] The question then is whether, in the face of significant unknowns with potentially devastating impacts on humanity, cost-benefit analysis has a role to play in the application of the precautionary principle to AI.

---

79. PEEL, *supra* note 22, at 4, 129.

80. For instance, California's Safe and Secure Innovation for Frontier Artificial Intelligence Models Act was proposed "to mitigate the risk of catastrophic harms from AI models so advanced that they *are not yet known to exist*" (Senate Third Reading, SB 1047, emphasis added).

81. Senate Rules Committee, SB 1047, Third Reading, https://leginfo.legislature.ca.gov/faces/billAnalysisClient.xhtml?bill_id=202320240SB1047.

82. For instance, for the California Chamber of Commerce, the Safe and Secure Innovation for Frontier Artificial Intelligence Models Act would "inevitably discourage[] economic and technological innovation," impact on "AI research and development in California," and make "AI business too risky in California, particularly given the potential penalties under SB 1047." Senate Third Reading, SB 1047; *see also*, John O. McGinnis, *The Folly of Regulating against AI's Existential Threat*, in THE CAMBRIDGE HANDBOOK OF ARTIFICIAL INTELLIGENCE: GLOBAL PERSPECTIVES ON LAW AND ETHICS 408–18 (Larry A. DiMatteo, Cristina Poncibò & Michel Cannarsa eds., 2022); Daniel Castro & Michael McLaughlin, *Ten Ways the Precautionary Principle Undermines Progress in Artificial Intelligence*, INFO. TECH. & INNOVATION FOUND. (Feb. 4, 2019), https://itif.org/publications/2019/02/04/ten-ways-precautionary-principle-undermines-progress-artificial-intelligence; John Bailey, *Treading Carefully: The Precautionary Principle in AI Development*, AM. ENTER. INST. (July 25, 2023) https://www.aei.org/technology-and-innovation/treading-carefully-the-precautionary-principle-in-ai-development; *but see* Alessio Tartaro, Adam Leon Smith & Patricia Shaw, *Assessing the impact of regulations and standards on innovation in the field of AI*, Cornell University, arXiv:2302.04110 [Preprint] (Feb. 8, 2023) https://arxiv.org/abs/2302.04110 (arguing that "in areas where they are considered high-risk, innovation will be supported by clear technical requirements, and the opportunity to directly discuss and test ideas with immediate regulatory feedback").

83. CASS R. SUNSTEIN, LAWS OF FEAR: BEYOND THE PRECAUTIONARY PRINCIPLE 14 (2005).

To answer this question under international law, it is first necessary to determine the application of the international principle of precaution to the AI existential threat. Although the precautionary principle first originated in relation to environmental protection, the principle has been extended to other domains such as public health and, critically for our purposes, the development of potentially dangerous technologies more generally. It is then necessary to examine whether the principle requires or merely justifies regulatory action to address potentially irreversible damage and what role a cost-benefit analysis should play in this assessment.

### A. *Relevance of the Precautionary Principle for AI Governance*

The precautionary principle, or approach, is based on the need for regulatory intervention to avoid potentially severe damage to human health or the environment despite scientific uncertainty on the relevant risks.[84] More specifically, "where there are threats of serious or irreversible damage, lack of full scientific certainty shall not be used as a reason for postponing cost-effective measures to prevent environmental degradation."[85] The principle thus translates into international law as the concept of precaution, or prudence, in response to risks.[86]

#### 1. The Precautionary Principle under International Law

The precautionary principle emerged in the 1980s from an increasing awareness of the risks caused by human activities and the realization of the limits of scientific analysis in the timely identification and management of uncertain risks.[87] As explained by International Court of Justice Judge Cançado Trindade, "the necessity to anticipate risks became increasingly manifest, as man continued to engage in evermore hazardous activities. Adopting a precautionary approach . . . became necessary to avoid environmental and health disasters."[88] The alternative approach, of postponing regulatory action until there is conclusive scientific proof of harm, has resulted in irreversible damage.[89] Similarly, many AI experts are now warning that rapid technological developments in AI present potentially significant dangers to humanity that cannot yet fully be understood based on current scientific analysis.[90] As explained by Wu, China's representative at the 2023 AI Safety Summit, AI technologies are "uncertain, unexplainable", and thus pose potentially significant risks for humanity that require government responses under international law.[91]

---

84. PEEL, *supra* note 22, at 112; *see also* DUPUY & VIÑUALES, *supra* note 24, at 70 (the underlying idea of precaution as a legal concept is that "the lack if scientific certainty about the actual or potential effects of an activity must not prevent States from taking appropriate measures when such effects may be serious or irreversible"); *see also* Antônio Augusto Cançado Trindade, *Principle 15*, *in* THE RIO DECLARATION ON ENVIRONMENT AND DEVELOPMENT: A COMMENTARY 403 (Jorge E. Viñuales ed., 2015).

85. Principle 15, U.N. Conference on Environment and Development, *Rio Declaration on Environment and Development*, U.N. Doc. A/CONF.151/26/Rev.1 (Vol. I), annex I (Aug. 12, 1992) [hereinafter Rio Declaration].

86. *See* JAMES CRAWFORD, BROWNLIE'S PRINCIPLES OF PUBLIC INTERNATIONAL LAW 341 (9th ed. 2019); PEEL, *supra* note 22, at 132.

87. Cançado Trindade, *supra* note 84, at 406.

88. *Id.* at 4.

Besides broad state practice on the application of precaution in the regulation of environmental risks,[92] an important number of international conventions on environmental protection recognize the precautionary principle, such as in the areas of climate change and transboundary air pollution,[93] the marine environment and transboundary watercourses,[94] and biodiversity.[95] On this basis, the precautionary principle is generally considered as part of the general principles of international environmental law.[96] The principle has also

---

89. PEEL, *supra* note 22, at 129.
90. *See* Part III.2, *infra*.
91. Cristina Criddle, Madhumita Murgia & Anna Gross, *US, China and 26 other nations agree to co-operate over AI development*, Financial Times (Nov. 1, 2023) https://www.ft.com/content/0869d0ec-a6fd-4fec-844a-61f837ed21a9.
92. SANDS ET AL., *supra* note 24, at 239–40; Kristel De Smedt & Ellen Vos, *The Application of the Precautionary Principle in the EU*, *in* THE RESPONSIBILITY OF SCIENCE 163–86 (Harald A. Mieg ed., 2022).
93. *See, e.g.*, Montreal Protocol on Substances that Deplete the Ozone Layer (with annex), pmbl., Jan. 1, 1989, 1522 U.N.T.S. 3, 26 I.L.M. 1541; United Nations Framework Convention on Climate Change, art. 3(3), May 9, 1992, S. Treaty Doc No. 102–38, 1771 U.N.T.S. 107 [hereinafter UNFCCC]; Convention on Long-Range Transboundary Air Pollution, Nov. 13, 1979, T.IA.S. 1054, 18 I.L.M. 1442.
94. Agreement for the Implementation of the Provisions of the United Nations Convention on the Law of the Sea of 10 December 1982 relating to the Conservation and Management of Straddling Fish Stocks and Highly Migratory Fish Stocks, Aug. 4, 1995, 2167 U.N.T.S. 3; Convention on the Prevention of Marine Pollution by Dumping Wastes and Other Matter, Dec. 29, 1972, 1046 U.N.T.S 120; Convention for the Protection of the Marine Environment of the North-East Atlantic [hereinafter OSPAR Convention] (opened for signature Sept. 22, 1992, entered into force Mar. 25, 1998); Convention on the Protection of the Marine Environment of the Baltic Sea Area, Mar. 22, 1974, 1507 U.N.T.S 166.
95. Convention on Biological Diversity, June 5, 1992, 1760 U.N.T.S. 79; Cartagena Protocol on Biosafety to the Convention on Biological Diversity, Jan. 29, 2000, 2226 U.N.T.S. 208 [hereinafter Cartagena Protocol].
96. *See, e.g.*, Responsibilities and Obligations of States sponsoring Persons and Entities with respect to Activities in the Area (Request for Advisory Opinion submitted to the Seabed Disputes Chamber), ITLOS Case No. 17, Advisory Opinion of Feb. 1, 2011, ITLOS (Seabed Dispute Chamber), ¶ 135 (referring to "a trend towards making this [i.e. the precautionary] approach part of customary international law"); *see also* SANDS ET AL., *supra* note 24, at 239–40 ("there is certainly sufficient evidence of state practice to support the conclusion that the principle . . . has now received sufficiently broad support to allow a strong argument to be made that it reflects a principle of customary law, and that within the context of the European Union it has now achieved customary status, without prejudice to the precise consequences of its application in any given case"); Cançado Trindade, *supra* note 84, at 413; Owen McIntyre and Thomas Mosedale, *The precautionary principle as a* found broader recognition and application in the area of public health, as reflected in international conventions on chemicals and biotechnology,[97] and the WTO Agreement on the Application of Sanitary and Phytosanitary Measures.[98] According to Sands and Peel, "the precautionary principle has now received widespread support by the international community in relation to a broad range of subject areas" and plays a "role in international law outside of the environmental field."[99] Peel confirms that there has been a "widespread adoption of the precautionary principle in international law."[100]

In sum, there is support for a broad interpretation of the precautionary principle under international law, not limited to addressing uncertain risks of environmental damage. Taking into account its application to public health and safety, the precautionary principle is relevant to address the potentially irreversible, but still uncertain, hazard posed by AI. The anthropocentric origins of the

---

*norm of customary international law* 9 J. ENVTL. L. 221, at 241 (1997) (arguing that the "precautionary principle has indeed crystallised into a norm of customary international law"); *but see* Panel Report, *European Communities — Measures Affecting the Approval and Marketing of Biotech Products*, ¶ 7.88, WTO Doc. WT/DS291/R (adopted Nov. 21, 2006) ("the legal debate over whether the precautionary principle constitutes a recognized principle of general or customary international law is still ongoing. Notably, there has, to date, been no authoritative decision by an international court or tribunal which recognizes the precautionary principle as a principle of general or customary international law").

97. Stockholm Convention on Persistent Organic Pollutants, pmbl. & art. 1, May 22, 2001, 2256 U.N.T.S. 119; 40 I.L.M. 532; Kyiv Protocol on Pollutant Release and Transfer Registers to the UNECE Convention on Access to Information, Public Participation in Decision-making and Access to Justice in Environmental Matters, May 21, 2003, 2629 U.N.T.S. 119; Cartagena Protocol, *supra* note 95, pmbl., arts. 1, 10(6), 11(8).

98. Annex 1A to the Marrakesh Agreement Establishing the World Trade Organization, art. 5.7, Apr. 15, 1994, 1867 U.N.T.S. 154, Agreement on the Application of Sanitary and Phytosanitary Measures, 1867 U.N.T.S. 493 ("In cases where relevant scientific evidence is insufficient, a Member may provisionally adopt sanitary or phytosanitary measures on the basis of available pertinent information, including that from the relevant international organizations as well as from sanitary or phytosanitary measures applied by other Members."). The General Court of the European Union extended the precautionary principle from environmental protection to health and consumer safety, and on this basis recognized precaution as a general principle of EU law. *See* Joined Cases T-74/00, 76/00, 83/00, 84/00, 85/00, 132/00, 137/00 & 141/00, Artegodan GmbH v. Comm'n, 2022 E.C.R. II-4945, ¶ ¶ 183–184 ("although the precautionary principle is mentioned in the Treaty only in connection with environmental policy, it is broader in scope. It is intended to be applied in order to ensure a high level of protection of health, consumer safety and the environment in all the Community's spheres of activity"); *see also* PAUL CRAIG, EU ADMINISTRATIVE LAW 697–698 (3rd ed. 2018).

99. SANDS ET AL., *supra* note 24, at 233–34; *but see* Arie Trouwborst, *The precautionary principle in general international law: combating the Babylonian confusion,* 16.2 RECIEL 185, at 190 (2007) (arguing that "in general international law as it stands, the reach of the precautionary principle is restricted to the environment"); *see also* Appellate Body Report, *European Communities — Measures Concerning Meat and Meat Products (Hormones)*, ¶ 123, WTO Doc. WT/DS26/AB/R (adopted Feb. 13, 1998) (considering that "the precautionary principle, at least outside the field of international environmental law, still awaits authoritative formulation").

100. PEEL, *supra* note 22, at 132.

principle further support its relevance for the protection of humanity from the existential risks posed by AI.

### 2. The Anthropocentric Origins of the Precautionary Principle and its Relevance to AI Safety

The 1992 Rio Declaration on Environment and Development, which provided one of the most influential formulations of the precautionary principle,[101] was based on the recognition that "human beings are at the centre of concerns for sustainable development. They are entitled to a healthy and productive life in harmony with nature."[102] The 1972 Stockholm Declaration on the Human Environment emphasized that "both aspects of man's environment, the natural and the man-made, are essential to his well-being and to the enjoyment of basic human rights—even the right to life itself."[103] In this manner, precaution originated as part of international efforts to protect human health and preserve human life.

As recognized in the Stockholm Declaration, international environmental protection was developed based on the recognition that "man's capability to transform his surroundings, if used wisely, can bring to all peoples the benefits of development and the opportunity to enhance the quality of life. Wrongly or heedlessly applied, the same power can do incalculable harm to human beings and the human environment."[104] Today, similar concerns characterize the development of AI technology, which promises great benefit but also generates significant concern regarding the potential threat it poses to human life. The principles of international environmental law that were developed to address the damage of human activities to the environment are now relevant to address the potential damage of AI to humanity.

### 3. Applying the Precautionary Principle to Address AI's Existential Risk

As a large number of leading AI experts are increasingly worried about the technology's potential threat to humanity, the level of alarm meets the threshold of "threats of serious or irreversible damage" that defines the application of the precautionary principle under international law.[105] While scientific opinion on whether these threats are likely to materialize is divided, "lack of full scientific certainty" is not a valid reason for not addressing these threats.[106] As many AI experts argue, waiting for conclusive scientific proof to address the potential existential risks posed by AI may mean a response proves to be too late.[107] This is similar to the argument made by small island states in support of using the precautionary principle in the climate change negotiations in the 1990s when the scientific consensus on the anthropogenic cause of climate change was less robust than today. These states argued that "we do not have the luxury of waiting for conclusive proof, as some have suggested in the past. The proof, we fear, will kill us."[108] As stated by the International Tribunal on the Law of the Sea ("ITLOS") in its Advisory Opinion on Climate Change and International Law, in determining necessary measures to prevent marine pollution from anthropogenic greenhouse gas emissions, "scientific certainty is not required."[109] In the absence of certainty, "States must apply the precautionary approach," which is "all the more necessary given the serious and irreversible damage that may be caused."[110]

---

101. Cançado Trindade, *supra* note 84, at 403.
102. Principle 1*5*, Rio Declaration, *supra* note 84. For Jorge Viñuales, "Principle 1 of the Rio Declaration takes a clearly anthropocentric stance structuring efforts towards sustainable development around human beings." *See* Jorge E. Viñuales, *The Rio Declaration on Environment and Development: Preliminary Study*, *in* THE RIO DECLARATION ON ENVIRONMENT AND DEVELOPMENT: A COMMENTARY, *supra* note 84, at 22.
103. Item 1, Stockholm Declaration, Report of the U.N. Conference on the Human Environment, U.N. Doc. A/CONF.48/ (June 16, 1972) 2.
104. *Id.* at Item 3.

### B. *Obligation to Regulate AI under the Precautionary Principle?*

Although there is relatively strong support for the recognition of the principle of precaution under international law, the exact meaning of the principle, and thus the obligations it imposes on states, remains unclear.[111] A key question for the international response to the potential threat that AI poses to humanity is whether the principle requires states to take action to address this threat or whether it merely justifies or guides state action.[112]

---

105. *See, e.g.*, Principle 15, Rio Declaration, *supra* note 85; UNFCCC, *supra* note 93, art. 3(3). The threat of AI has been compared to climate change; *see, e.g.*, Martin Coulter, *AI pioneer says its threat to the world may be 'more urgent' than climate change*, REUTERS (May 9, 2023), https://www.reuters.com/technology/ai-pioneer-says-its-threat-world-may-be-more-urgent-than-climate-change-2023-05-05.

106. Principle 15, Rio Declaration, *supra* note 85; s*ee also* Convention on Biological Diversity, *supra* note 95, pmbl.; UNFCCC, *supra* note 93, art. 3(3).

107. *See, e.g.*, RUSSELL, *supra* note 34, at 151 ("[I]f we consider the global catastrophic risks from climate change, which are predicted to occur later in this century, is it too soon to take action to prevent them? On the contrary, it may be too late. The relevant time scale for superhuman AI is less predictable, but of course that means it, like nuclear fission, might arrive considerably sooner than expected"); *see also* Michael Shermer, *Artificial Intelligence Is Not a Threat—Yet*, SCI. AM. (Mar. 1, 2017), https://www.scientificamerican.com/article/artificial-intelligence-is-not-a-threat-mdash-yet ("Yudkowsky thinks that if we don't get on top of this now it will be too late: 'The AI runs on a different timescale than you do; by the time your neurons finish thinking the words 'I should do something' you have already lost.'").

108. Ambassador Robert van Lierop, Permanent Representative of Vanuatu to the U.N. and Co-Chairman of Working Group 1 of the Intergovernmental Negotiating Committee for a Framework Convention on Climate Change (INF/FCCC), Statement to the Plenary Session of the INC/FCCC (Feb. 5, 1991), at 3, *cited in* SANDS ET AL., *supra* note 24, at 230.

109. Request for an Advisory Opinion submitted by the Commission of Small Island States on Climate Change and International Law (Request for Advisory Opinion submitted to the Tribunal) ITLOS Case No. 31, Advisory Opinion of May 21, 2024, ¶ 213.

110. *Id.*

### 1. Addressing Potential Risks

The Rio Declaration uses mandatory language in its formulation of the precautionary approach.[113] In case of serious threats, scientific uncertainty "shall not be used as a reason for postponing" action to prevent the potential damage.[114] Judge Cançado Trindade goes one step further by arguing that the precautionary principle could be understood "not only as a justification not to postpone action but potentially as an obligation to take action despite the absence of conclusive evidence regarding the harm an activity may cause. . . ."[115] Under this broad interpretation, the precautionary principle imposes on states an obligation of prudence and due diligence.[116]

In its Advisory Opinion on *Responsibilities in the Area*, the ITLOS Seabed Dispute Chamber linked the precautionary approach to states' general obligation of due diligence and, on this basis, found that states were "required . . . to take all appropriate measures to prevent damage that might result from the activities of contractors that they sponsor."[117] For ITLOS, this due diligence and precautionary obligation "applies in situations where scientific evidence concerning the scope and potential negative impact of the activity in question is insufficient but where there are plausible indications of potential risks."[118] Disregarding these potential risks would fail to meet the obligation of due diligence and amount to a breach of the precautionary approach.[119] In *Tătar v. Romania*, the European Court for Human Rights emphasized the importance of the precautionary principle, as formulated under the Rio Declaration, in support of its conclusion that Romania violated the right to private and family life under the European Convention for Human Rights by failing to seriously evaluate certain industrial risks and inform the general public.[120] The Courts of the European Union have also indicated that the precautionary principle entails an obligation on states to take regulatory action to address threats despite the absence of scientific certainty.[121] In particular, authorities must identify the potentially negative impact on the health and safety of new products, engage in a thorough assessment of risks based on the most relevant and recent research, and take action where there is still scientific uncertainty on the existence or extent of these risks.[122] The state must act "as soon as it [becomes] aware of serious scientific information alleging the existence of potential risks to human health to which a relatively new product on the market might give rise."[123]

---

111. *See, e.g.*, SANDS ET AL., *supra* note 24, at 234 ("there is no clear and uniform understanding of the meaning of the precautionary principle among states and other members of the international community").
112. On justification versus obligation under the principle of precaution, see Cançado Trindade, *supra* note 84, at 407–08.
113. SANDS ET AL, *supra* note 24, at 234.
114. Principle 15, Rio Declaration, *supra* note85.
115. Cançado Trindade, *supra* note 84, at 408; *see also, e.g.*, Trouwborst, *supra* note 99, at 188 ("Wherever, on the basis of the best information available, there are reasonable grounds for concern that serious and/or irreversible harm to the environment may be caused, effective and proportional action to prevent and/or abate this harm *must be taken*, including in the face of scientific uncertainty regarding the cause, extent and/or probability of the potential harm.") (emphasis added).
116. Cançado Trindade, *supra* note 84, at 409. On precaution and due diligence, see also DANIEL BODANSKY, JUTTA BRUNNÉE & LAVANYA RAJAMANI, INTERNATIONAL CLIMATE CHANGE LAW 43–44 (2017); BENOIT MAYER, THE INTERNATIONAL LAW ON CLIMATE CHANGE 71 (2018) ("The growing recognition of a precautionary principle . . . suggests that some sovereign obligations arise as soon as credible 'threats of serious or irreversible damage' are perceived"); BENOIT MAYER, INTERNATIONAL LAW OBLIGATIONS ON CLIMATE CHANGE MITIGATION 99 (2022).
117. ITLOS Case No. 17, *supra* note 96, ¶ 131; *see also* Southern Bluefin Tuna (New Zealand v. Japan; Australia v. Japan), Provisional Measures, ITLOS Cases Nos. 3 & 4, Order of Aug. 27, 1999, ITLOS Rep. 1999, at 274, ¶ 77; ITLOS Case No. 31, *supra* note 109, ¶ 242 (confirming the close link between states' obligation of due diligence and the precautionary approach, and in particular states' obligation to "apply the precautionary approach in their exercise of due diligence to prevent, reduce and control marine pollution from anthropogenic GHG emissions").

### 2. The Precautionary Principle, International Cooperation, and Irreversible Threats

At the international level, a requirement to act with caution can be interpreted as an obligation of states to cooperate in addressing potentially irreversible threats.[124] ITLOS, for instance, considered in the *MOX Plant Case* that "prudence and caution" require states to cooperate in exchanging information concerning risks and devising ways to deal with those risks.[125] More generally, cooperation based on precaution is part of a number of international treaties on environmental protection in cases where the environmental issue at stake remains characterized by scientific uncertainty.[126] On a similar basis, precautionary action in addressing the threat posed by AI technologies would require international cooperation in adopting measures to anticipate, prevent, or minimize the causes of these threats and "mitigate [their] adverse effects."[127]

---

118. ITLOS Case No. 17, *supra* note 96, ¶ 131.
119. *Id.*
120. Tătar c. Roumanie, App. No. 67021/01, 115 Eur. Ct. H.R. (Jan. 27, 2009, final version issued July 6, 2009) (in French) ¶ 120.
121. For instance, Artegodan GmbH v. Comm'n, ¶ 184 (interpreting the precautionary principle as "*requiring* the competent authorities to take appropriate measures to prevent potential risks to public health, safety and the environment …") (emphasis added); *see also* Case T-392/02, Solvay Pharmaceuticals BV v Council of the European Union, 2003, E.C.R. II-04555, ECLI:EU:T:2003:277, ¶ 121 (Oct. 21, 2003); Case C-477/14, Pillbox 38 (UK) Ltd v Secretary of State, ECLI:EU:C:2016:324, ¶ 116 (May 4, 2016) ("which *require* it to ensure a high level of protection of human health in the definition and implementation of all Union policies and activities") (emphasis added).
122. *See* CRAIG, *supra* note 98, at 699 (summarizing the relevant case law of the European judiciary on the application of the precautionary principle).
123. Pillbox 38 (UK) Ltd v Secretary of State, ¶ 116.
124. *See* The MOX Plant Case (Ireland v U.K.), Provisional Measures, ITLOS Case No. 10, Order of Dec. 3, 2001, ¶ 84 (stating that "prudence and caution require that Ireland and the United Kingdom cooperate in exchanging information concerning risks or effects of the operation of the MOX plant and in devising ways to deal with them, as appropriate").

Although this interpretation has not yet received sufficient international support, the precautionary principle could also in theory operate as a reversal of the burden of proof, from the entity opposing the conduct of a potentially hazardous activity to the one that wishes to carry it out.[128] Following this approach, the mere presumption of serious and irreversible damage would be sufficient to require the adoption of precautionary measures.[129] Accordingly, states should not authorize the placement of potentially harmful products on the market unless those offering these products can prove that they are safe.[130] According to the U.N. General Assembly World Charter for Nature, "where potential adverse effects are not fully understood, the activities should not proceed."[131] Under the 1992 Convention for the Protection of the Marine Environment of the North-East Atlantic, the states that wanted to retain the option of dumping certain radioactive wastes at sea had to report "the results of scientific studies which show that any potential dumping operations would not result in hazards to human health, harm to living resources or marine ecosystems . . . ."[132] As explained by Judge Weeramantry in his Dissenting opinion regarding New Zealand's request for the continuation of the *Nuclear Tests* case proceedings, reversing the burden of proof in the case of the threat of possible irreversible damage is logical given that the necessary information on that threat may largely be in the hands of the party causing or threatening the damage.[133] Similarly, it is logical to transfer the burden of proof to AI developers, as they should fully understand the threat posed by their products before placing them on the market. As the 'risks associated with AI systems can only be effectively addressed at the design stage', the placement of AI products on the market should not be authorized until the potential adverse effects of these technologies are fully understood, including their potential existential threat.[134]

---

125. ITLOS Case No. 10, Order of Dec. 3, 2001. On "prudence and caution," see also ITLOS Cases Nos. 3 & 4, *supra* note 117, ¶ 77.

126. *See, e.g.*, UNFCCC, *supra* note 93, art. 3(3) (noting that "the Parties should take precautionary measures to anticipate, prevent or minimize the causes of climate change and mitigate its adverse effects. . . . Efforts to address climate change may be carried out cooperatively by interested Parties").

127. *Id.* On the importance of international cooperation to address AI security risks, see also *Council of Europe Framework Convention on Artificial Intelligence, supra* note 15, at Article 25; *OECD AI Principles, supra* note 15, item 2.5.

128. SANDS ET AL., *supra* note 24, at 234. In Pulp Mills on the River Uruguay (Arg. v. Uru.), Judgment, 2010 I.C.J. Rep. 14, ¶ 164 (Apr. 20), the ICJ considered "that while a precautionary approach may be relevant in the interpretation and application of the provisions of the Statute, it does not follow that it operates as a reversal of the burden of proof."

129. *See* Int'l L. Association, *Committee on the Legal Principles Relating to Climate Change, Sofia Conference, Second Report*, at 30 (Aug. 26, 2012) (". . .there is some State practice to indicate that a form of burden shifting may be required by the precautionary principle, in that sense that the mere presumption of irreversible harm may be sufficient to trigger precautionary measures"); *see also* Cançado Trindade, *supra* note 84, at 409 ("the precautionary approach, when understood as a broad obligation of prudence, finds an implicit place in provisions regarding the protection of ecosystems and in the obligation of due diligence").

130. SANDS ET AL., *supra* note 24, at 234.

131. G.A. Res. 37/7, annex, art. 11(b) (Oct. 28, 1982).

132. OSPAR Convention, *supra* note 94, at 23.

An alternative and weaker interpretation of the precautionary principle is that it does not require but rather justifies restrictive regulatory measures. According to Crawford, the principle "can be interpreted to imply that precautionary regulation is justified when there is no clear evidence about a particular risk scenario, when the risk itself is uncertain, or until the risk is disproved."[135] This interpretation can help shield states from claims under international and regional economic law. While the principle does not relieve states of their international trade and investment obligations,[136] the absence of scientific certainty to adopt protective measures does not prevent

---

133. Request for an Examination of the Situation in Accordance with Paragraph 63 of the Courts Judgment of 20 December 1974 in the Nuclear Tests Case (N.Z. v. Fr.), Dissenting Opinion of Judge Weeramantry, 1995 I.C.J. 348, 348 (Sept. 2).

134. *See Explanatory Report to the Council of Europe Framework Convention on Artificial Intelligence, supra* note 15, at item 90 ("It is also important to recognise that some artificial intelligence developers, including those with a public interest mission, cannot proceed with their innovation unless they can be reasonably sure that it will not have harmful implications and incorporate appropriate safeguards to mitigate risks in a controlled environment"); *Council of Europe Framework Convention on Artificial Intelligence, supra* note 15, at Article 16(4) ("Each Party shall assess the need for a moratorium or ban or other appropriate measures in respect of certain uses of artificial intelligence systems where it considers such uses incompatible with the respect for human rights, the functioning of democracy or the rule of law"). In this sense, see also SB-1047, 2023-2024 Leg., Reg. Sess. (Cal. 2024), at 22603(b) (stating before making their models available for commercial or public use, the developers of covered AI models shall "assess whether the covered model is reasonably capable of causing or materially enabling a critical harm" and "take reasonable care to implement appropriate safeguards to prevent the covered model and covered model derivatives from causing or materially enabling a critical harm"). Similarly, *see EU Artificial Intelligence Act, supra* note 14, Article 5 (prohibiting the "placing on the market" of certain AI systems).

135. CRAWFORD, *supra* note 86, at 341.

136. *See European Communities — Measures Concerning Meat and Meat Products*, *supra* note 99, ¶ 124 ("however, the precautionary principle does not, by itself, and without a clear textual directive to that effect, relieve a panel from the duty of applying the normal (i.e. customary international law) principles of treaty interpretation in reading the provisions of the SPS Agreement").

states from justifying restrictive measures taken to protect human health and safety.

For instance, in *European Communities — Asbestos*, the WTO Appellate Body ruled that, in justifying state measures under the 1994 General Agreement on Trade and Tariffs, "a Member may also rely, in good faith, on scientific sources which, at that time, may represent a divergent, but qualified and respected, opinion. A Member is not obliged, in setting health policy, automatically to follow what, at a given time, may constitute a majority scientific opinion."[137] In *European Communities — Hormones*, it is stated that "responsible and representative governments may act in good faith on the basis of what, at a given time, may be a divergent opinion coming from qualified and respected sources."[138] Similarly, the Court of Justice of the European Union has accepted that the "precautionary principle justifies the adoption of restrictive measures . . . where it proves impossible to determine with certainty the existence or the extent of the risk envisaged because of the insufficiency, inconclusiveness or imprecision of the results of the studies concerned."[139] In case of scientific uncertainty, "protective measures may be taken without having to wait until the reality and seriousness of those risks become fully apparent,"[140] allowing states to derogate from requirements on the free movement of goods.[141] Applied to AI safety measures, the

---

137. Appellate Body Report, *European Communities — Measures Affecting Asbestos and Products Containing Asbestos*, ¶ 178, WTO Doc. WT/DS135/AB/R (adopted Apr. 5, 2001).

138 *European Communities — Measures Concerning Meat and Meat Products*, *supra* note 99, ¶ 194. This is particularly important in cases "where risks of irreversible, e.g. life-terminating, damage to human health are concerned." *European Communities — Measures Concerning Meat and Meat Products*, *supra* note 99, ¶ 124.

139. Case C-219/07, Nationale Raad van Dierenkwekers en Liefhebbers VZW and Andibel VZW v. Belgische Staat, 2008 E.C.R. I-4477, ¶ 38; *see also, e.g.*, Case C-77/09, Gowan Comércio Internacional e Serviços Lda v. Ministero della Salute, 2010 E.C.R. I-13555, ¶ 76,

> Where it proves to be impossible to determine with certainty the existence or extent of the alleged risk because of the insufficiency, inconclusiveness or imprecision of the results of studies conducted, but the likelihood of real harm to public health persists should the risk materialise, the precautionary principle justifies the adoption of restrictive measures, provided they are non-discriminatory and objective.

140. Case C-236/01, Monsanto Agricoltura Italia SpA v. Presidenza del Consiglio dei Ministri, 2003 E.C.R. I-8166, ¶ 111; *see also* Case C-180/96, United Kingdom v. Comm'n, 1998 E.C.R. I-2269, ¶ 99; Case C-157/96, National Farmers' Union and Others, 1998 E.C.R. I-2236, ¶ 63; Case E-3/00, EFTA Surveillance Auth. v. The Kingdom of Norway, EFTA Ct. Rep. 73, Judgment of the Court, ¶ 31 (Apr. 5, 2001).

141. *See, e.g.*, Case C-473/98, Kemikalieinspektionen v. Toolex Alpha AB, 2000 E.C.R. I-5702, ¶ 45,

> The latest medical research on the subject, and also the difficulty of establishing the threshold above which exposure to trichloroethylene poses a serious health risk to humans, given the present state of the research, there is no evidence in this case to justify a conclusion by the Court that national legislation such as that at precautionary principle would thus help states justify restrictions adopted to protect the commercialization of AI products that could cause potentially irreversible damage.

### C. *Precaution and Cost-Benefit Analysis*

A common argument against regulating AI in response to its potential existential threat is that the concrete and predictable benefits of AI outweigh the speculative and remote risks of harm. For instance, invoking this kind of cost-benefit analysis, McGinnis argues that "there is an overwhelming case against the current regulation of AI for existential risks."[142] More specifically, AI regulation "will get in the way of continued progress in AI that delivers extraordinary benefits. . . . In contrast to the certainty of substantial benefits there is no consensus that we face existential risks from AI."[143] Similarly, Castro and McLaughlin warn that "if policymakers apply the 'precautionary principle' to AI . . . they will limit innovation and discourage adoption-undermining economic growth, competitive advantage, and social progress."[144] They argue that:

> too often policies based on the precautionary principle fail to strike the balance between addressing actual harms posed by AI and not hindering innovation. . . . These policies are misguided not because they create regulation, but because they create unnecessary barriers to developing and adopting AI due to exaggerated fears of AI or failures to recognize that existing or more nuanced regulation would address potential issues.[145]

According to critics of AI regulation, the potential threat of AI is not imminent and must be discounted by the unlikelihood that it will materialize. On this basis, "any regulation focused on this threat should be postponed to a time when regulators would enjoy greater knowledge."[146] This argument against AI regulation builds on previous criticism of the precautionary principle as impeding technological

---

issue in the case in the main proceedings goes beyond what is necessary to achieve the objective in view;

*see also* Joanne Scott, *The Precautionary Principle before the European Courts, in* PRINCIPLES OF EUROPEAN ENVIRONMENTAL LAW (PROCEEDINGS OF THE AVOSETTA GROUP OF EUROPEAN ENVIRONMENTAL LAWYERS) 62 (Richard Macrory, Ian Havercroft & Ray Purdy eds., 2004) (explaining how the precautionary principle "liberates" the European Union to take regulatory measures in the absence of scientific certainty).

142. McGinnis, *supra* note 82, at 418.
143. *Id.* at 410, 414.
144. Castro & McLaughlin, *supra* note 82, at 1. Similarly, *see Executive Order on Removing Barriers to American Leadership in Artificial Intelligence, supra* note 11, at Section 1 (revoking "certain existing AI policies and directives that act as barriers to American AI innovation, clearing a path for the United States to act decisively to retain global leadership in artificial intelligence").
145. *Id.* at 5.
146. McGinnis, *supra* note 82, at 414.

development, most notably Cass Sunstein's influential *Laws of Fear.*[147] It is also associated with concerns that a strong precautionary approach conceals a protectionist and anti-technological agenda.[148]

Cost-effectiveness plays a role in how the precautionary principle is interpreted under international law. According to the Rio Declaration, scientific uncertainty shall not lead states to postpone "cost-effective measures" to address irreversible damage.[149] Similarly, the reference to precaution in the United Nations Framework Convention on Climate Change refers to "cost-effective [measures] so as to ensure global benefits at the lowest possible cost."[150] However, cost-effectiveness cannot be equated with cost-benefit analysis.[151] Sunstein himself does not advocate a cost-benefit analysis in cases where there is a risk of catastrophic damage, but merely argues that cost-effective responses be employed (i.e., "regulators [choosing] the least costly means of achieving their ends").[152] Sunstein accepts that in cases of potential catastrophic risk, for which the likelihood cannot be determined through probabilities, "a large margin of safety makes a great deal of sense," and regulators do not need as much evidence that the risk of damage is probable.[153] Contrary to the criticism of the precautionary regulation of AI existential risks that relies on Sunstein's work, Sunstein agrees that "the Precautionary Principle has a legitimate place when people face a potentially catastrophic risk to which probabilities cannot be assigned."[154] On this basis, Sunstein advocates applying an "Anti-Catastrophe Principle" in regulatory policy.[155] Given the catastrophic consequences in a worst-case scenario regarding AI development, Sunstein's Anti-Catastrophe Principle would justify regulatory intervention to address the existential risk posed by AI, even in the absence of conclusive evidence regarding this threat.

---

147. *Id.* at 409 (referring to Sunstein's Laws of Fear, argues that "the strong form of the precautionary principle has been rightly criticized because it does not sufficiently consider the benefits of innovation that the regulation will prevent. Why should these be discounted more than the risks of harm? Doing so creates obstacles to progress and may create harm itself.").

148. *See, e.g.*, Nigel Cory & Patrick Grady, *The EU's Approach to AI Standards Is Protectionist and Will Undermine Its AI Ambitions*, CENTRE FOR DATA INNOVATION (Feb. 6, 2023), https://datainnovation.org/2023/02/the-eus-approach-to-ai-standards-is-protectionist-and-will-undermine-its-ai-ambitions ("if the EU wants to be a leader in AI, it needs to create standards that foster innovation, not hinder it"). More generally, on concerns "that allegations of uncertainty might be used to conceal a fervently anti-technological agenda," see PEEL, *supra* note 22, at 151. Precautionary regulation has also been criticized as an instrument of trade protectionism. *See, e.g.*, PEEL, *supra* note 22, at 117; Giandomenico Majone, *What Price Safety? The Precautionary Principle and its Policy Implications*, 40 J. COMM. MARK. STUD. 89, 89 (2002).

149. Rio Declaration, *supra* note 85, at Principle 15.

150. UNFCCC, *supra* note 93, art. 3(3).

151. *See, e.g., Communication from the Commission on the Precautionary Principle*, at 17–18, COM (2000) 1 final (Feb. 2, 2000) (according to which the proportionality of regulations to achieve the desired level of protection "cannot be reduced to an economic cost-benefit analysis"). In Artegodan GmbH v. Comm'n, 2022 E.C.R. II-4945, ¶ 184, the General Court of the EU interpreted precaution as requiring authorities to regulate the "potential risks to public health, safety and the environment, by giving precedence to the requirements related to the protection of those interests over economic interests."

152. SUNSTEIN, *supra* note 83, at 115. For instance, regarding climate change, "there are many methods by which to reduce the relevant risks. Both nations and international institutions should choose those methods that minimize costs." *Id.*

153. *Id.* at 116–17.

Having established that the precautionary principle may be applied to AI risk regulation, the following section examines how the principle may be applied under international human rights law.

### IV. THE EXISTENTIAL THREAT OF ARTIFICIAL INTELLIGENCE UNDER INTERNATIONAL HUMAN RIGHTS LAW

The previous section argued that the precautionary principle is established as a matter of international law and imposes obligations on states to act where there are potentially severe risks to human health despite the absence of conclusive evidence establishing the probability of the risk. It was noted that these obligations are varied and include a duty to regulate and to act prudently and with due diligence. It includes a duty to cooperate with other states to address the threat and provides legal justification in the event that state restrictions on AI development are challenged as contrary to international law.

International human rights law provides a useful avenue to advance the precautionary principle. In contrast to precaution as a principle applicable in other areas of international law, where it might be problematic to find an international tribunal with jurisdiction to test its scope in relation to the AI threat, numerous international institutions already possess the power to monitor human rights, together with international tribunals with the jurisdiction to entertain human rights complaints. Much like how climate change strategic litigations have turned to supranational human rights tribunals to give effect to the precautionary principle, this section argues that there is equal potential for litigants to test the legality of AI within the rubric of international human rights law.[156]

Building on the precautionary principle outlined in the previous section, the following section will show that this principle has taken hold within international human rights law, thereby providing additional impetus—as well as a viable campaigning and litigation option for concerned actors—to test the extent of the obligation imposed on states to address AI as an existential risk.

---

154. *Id.* at 225.
155. *Id.*
156. *See e.g.*, Therese Karlsson Niska, *Climate Change Litigation and the European Court of Human Rights - A Strategic Next Step?*, 13(4) J. WORLD ENERGY L. & BUS. 331 (2020) (discussing using the European Court of Human Rights for climate change litigation in the AI context). As to the nature of strategic litigation, see Michael Ramsden & Kris Gledhill, *Defining Strategic Litigation*, 4 CIV. JUST. Q. 407 (2019) (discussing the nature of strategic litigation).

### A.     *The Right to Life and AI's Existential Threat*

Although the development and use of AI has implications for a variety of human rights, in the context of existential threats to life the starting point here is to approach this from the perspective of the right to life. While the loss of human life ultimately implicates the enjoyment of all human rights—including the right to privacy and freedom of expression—the right to life is ultimately the core human right from which others flow.[157] Focussing on the right to life where life has been deprived is thus a logical foundation for the analysis of AI as an existential risk to life.

In this regard, the right to life is provided at the international level in the International Covenant on Civil and Political Rights ("ICCPR"), and at the regional level in various instruments, including the European Convention on Human Rights ("ECHR") and the American Convention on Human Rights ("ACHR").[158] The universal recognition of this right by states also establishes its status as customary international law and, thus, a robust source of international obligation irrespective of whether a state is a party to a treaty that contains this right.[159]

The various treaty bodies that monitor the implementation of the right to life have begun to address the implications of AI, but not all in negative terms.[160] In fact, monitors have discussed the value of AI in realizing the right to life and advancing health and social well-being.[161] The UNESCO Recommendation on the Ethics of AI thus expressly encouraged states to "employ effective AI systems for improving human health and protecting the right to life, including mitigating disease outbreaks, while building and maintaining international solidarity to tackle global health risks and uncertainties."[162] The use of AI is therefore not only permitted by international human rights law, but might also be a preferred method to enhance rights in some instances.

---

157. Human Rights Comm., General Comment 36 on Article 6: right to life, ¶ 2, U.N. Doc. CCPR/C/GC/36 (Sep. 3, 2019) [hereafter General Comment 36]. See also *Council of Europe Framework Convention on Artificial Intelligence, supra* note 15, art 1 (noting the general requirement that "the lifecycle of artificial intelligence systems are fully consistent with human rights").

158. International Covenant on Civil and Political Rights art. 6, Dec. 16, 1966, 999 U.N.T.S. 171 [hereinafter ICCPR]; European Convention for the Protection of Human Rights and Fundamental Freedoms art. 2, Nov. 4, 1953, 213 U.N.T.S. 221 [hereinafter ECHR]; American Convention on Human Rights art. 4, Nov. 22, 1978, 1144 U.N.T.S. 123; *see also* African Charter on Human and Peoples' Rights art. 4, June 27, 1986, 1520 U.N.T.S. 217.

159. *See* Stuart Casey-Maslen, *Customary Rules Pertaining to the Right to Life* 735, 735, *in* THE RIGHT TO LIFE UNDER INTERNATIONAL LAW: AN INTERPRETATIVE MANUAL (2021).

160. As to the impact on other rights, including the right to privacy and right to equality, see, e.g., McGregor et al., *supra* note 18, at 310 (noting that the use of AI in automated credit scores can affect employment and housing rights; the use of algorithms to inform social security decision-making potentially interferes with a range of social rights; and their use to assist with identifying children at risk may interference with the right to family life).

161. UNESCO, RECOMMENDATION ON THE ETHICS OF AI 37 (2021).

Yet, as will be argued here, the right to life also contains an obligation upon states to assess the threat posed by AI and take steps to address this threat. Accordingly, the following section considers the implications of the right to life for AI development and use. It is structured in two parts. The first part considers the right to life as a negative right in prohibiting arbitrary deprivations of life. This negative prohibition has a clear application to any future uses of AI that lead to the loss of human life but also is limited in scope to only state agents engaged in AI design and use. This leads to the second aspect that will be considered: the extent to which the right to life triggers a positive obligation to prevent future deprivations of life, which should include measures that cover both state and private enterprises alike.[163]

### B.    *The Prohibition on Taking Life and Its Applicability to AI*

According to its most generally accepted definition, the right to life expressly protects individuals from "arbitrary" deprivations of life.[164] The right is thus not absolute as such; an interference should not be arbitrary.[165] Typically, this would apply to those instances where lethal force is used to prevent an imminent threat to human life.[166] Similarly, the use of lethal force during armed conflict that complies with international humanitarian law would also not constitute an arbitrary deprivation of life.[167] In these situations, to the extent that AI complies with these requirements, there would not be a violation of the right to life. However, more generally, the Human Rights Committee, the body in charge of monitoring compliance with the ICCPR, has defined an arbitrary deprivation of life as involving an interference that is (1) not prescribed by law, (2) disproportionate to the ends sought, and (3) not necessary in the circumstances of the case considering the availability of less harmful means.[168]

---

162. *Id.*; *see also An Open Letter: Research Priorities for Robust and Beneficial Artificial Intelligence*, Oct. 28, 2015, FUTURE OF LIFE INSTITUTE, http://futureoflife.org/ai-open-letter (noting that "everything that civilization has to offer is a product of human intelligence; we cannot predict what we might achieve when this intelligence is magnified by the tools Al may provide, but the eradication of disease and poverty are not unfathomable.").

163. As to the general nature of positive obligations under IHRL, see Human Rights Comm., General Comment No. 31: The Nature of the Legal Obligation Imposed on States Parties to the Covenant, ¶¶ 3–8, U.N. Doc. CCPR/C/21/Rev.1/Add. 13 (May 26, 2004); UN Comm. on Economic Social and Cultural Rights, General Comment No. 3 The Nature of States Parties' Obligations, ¶¶ 2–8 UN Doc E/1991/23 (Dec. 14, 1990).

164. ICCPR, *supra* note 158, art. 6. There are slight variations in other instruments. For example, Article 2 of the ECHR, prohibits "intentional" deprivations of life "save in the execution of a sentence of a court following his conviction of a crime for which this penalty is provided by law". ECHR, *supra* note 158, art. 2. Where life is deprived in a law enforcement context, then the use of force must be "no more than absolutely necessary." *Id.* Despite the differing terminology, it is often considered that these standards are analogous.

165. General Comment 36, *supra* note 157, ¶ 10.

166. *See* Michael Ramsden, *Targeted Killings and International Human Rights Law: The Case of Anwar Al-Awlaki*, 16 J. CONFLICT & SEC. L. 385 (2011) (providing an example of a time in which a loss of life/violation of the right to life might be justified under IHRL).

The central question here, therefore, concerns whether, in the scenario in which the existential threat played itself out, the deprivation of life caused by AI was proportionate to a legitimate aim. Even if AI development is pursued for legitimate societal objectives (e.g., scientific or medical research or national defense), it is important to acknowledge that such goals could never be justifiable if the unintended consequence were humanity losing control over the technology and bringing about its own extinction. The inherent disproportionality of such an outcome supports the need for the state to proactively adopt precautionary measures, as the next section establishes.

### 1. Prioritizing the Right to Life in AI Design

The negative obligation not to arbitrarily deprive human life itself raises the question of whether AI can be programmed in a manner that complies with this right. AI design might do this by incorporating compliance with the right to life into the system's decision-making, drawing upon the significant body of caselaw and commentary on human rights.[169] An important point here is that human rights should be prioritized so that AI systems should never be allowed to violate human rights in the pursuit of their goals.[170] In this regard, as the

---

   167. *See also* African Commission on Human and Peoples' Rights, General Comment No. 3 on the African Charter on Human and Peoples' Rights: The Right to Life (Article 4), ¶¶ 12–13 (Nov. 18, 2015).
   168. Teitiota v New Zealand, CCPR/C/127/D/2728/2016, Human Rights Comm., Decision, ¶ 2.9 (Jan. 7, 2020); Gorji-Dinka v. Cameroon, CCPR/C/83/D/1134/2002, Human Rights Comm., Decision, ¶ 5.1 (Mar. 14, 2002); Van Alphen v. Netherlands, CCPR/C/39/D/305/1988, Human Rights Comm., Decision, ¶ 5.8 (July 23, 1990); Camargo v. Colombia, CCPR/C/OP/1, Human Rights Comm., Decision, ¶ 13.2 (Mar. 31, 1982); Eighth United Nations Congress on the Prevention of Crime and the Treatment of Offenders, *Basic Principles on the Use of Force and Firearms by Law Enforcement Officials*, ¶ 9 (Sept. 7, 1990); McCann v. United Kingdom, App. No. 18984/91, ¶ 150 (Sept. 27, 1995), https://hudoc.echr.coe.int/fre?i=001-57943.
   169. *See, e.g.*, Thomas Burri, *International Law and Artificial Intelligence*, 60 GER. Y.B. INT'L L. 91, 99 (2017) (emphasizing that autonomous weapons systems that "select and engage targets without meaningful human control" are more likely to be banned than weapons equipped with a degree of human oversight).
   170. Ondrej Bajgar & Jan Horenovsky, *Negative Human Rights as a Basis for Long-term AI Safety and Regulation*, 76 J. A.I. RSCH. 1043, 1053 (2023); *but also see* Hin-Yan Liu, *AI*

European Commission has observed, "[r]espect for fundamental rights, within a framework of democracy and the rule of law, provides the most promising foundations for identifying abstract ethical principles and values, which can be operationalised in the context of AI."[171] Whether and the extent to which technical solutions exist to this issue cannot be fully explored here; as noted in Part II with respect to the alignment problem, it remains unclear whether a solution is even technologically possible. Instead, it suffices to note that the obligation to not arbitrarily deprive humans of their life compels the design of AI to comply with this right. If this is not possible, then states have to consider more fundamentally the compatibility of ongoing AI programs with the right to life.[172]

### 2. State vs. Non-State Actors

Setting aside this technical question, a more general problem emerges concerning the limited scope of the right to life, which places obligations solely on state actors. This reflects the nature of international human rights law as rooted in a vertical relationship between the citizen and state, with the state considered the predominant oppressive power.[173] It is thus states that owe human rights obligations and who can be held responsible for violating human rights, for example, by depriving someone of their life without lawful justification.[174] As things currently stand, private actors are therefore under no legal obligation, at least under international human rights

---

*Challenges and the Inadequacy of Human Rights Protections*, 40(1) CRIM. JUST. ETHICS 2 (arguing that the monopoly of human rights on assessing human harm needs to be broken so that a greater range of barriers can be deployed).

171. European Commission Independent High-Level Expert Group on Artificial Intelligence, *Ethics Guidelines for Trustworthy AI* (2019), https://www.europarl.europa.eu/cmsdata/196377/AI%20HLEG_Ethics%20Guidelines%20for%20Trustworthy%20AI.pdf.

172. *See* Human Rights Watch, *Stopping Killer Robots: Country Positions on Banning Fully Autonomous Weapons and Retaining Human Control* (Aug. 10, 2020), at 1 (discussing the more specific context of autonomous weapons); Burri, *supra* note 169; Mathias Risse, *Human Rights and Artificial Intelligence: An Urgently Needed Agenda*, 41(1) HUM. RTS. Q. 1, at 10 (2019).

173. Kuzi Charamba, *Beyond the Corporation Responsibility to Protect Human Rights in the Dawn of a Metaverse*, 30(1) UNIV. MIAMI INT'L & COMP. L. REV. 110 (2022). Imposing direct human rights obligation on corporations continues to be a challenge. G.A., *Report of the Working Group on the Issue of Human Rights and Transnational Corporations and Other Business Enterprises on Access to Effective Remedies Under the Guiding Principles on Business and Human Rights: Implementing the United Nations Protect, Respect and Remedy Framework*, U.N. Doc. A/72/162, ¶ 5 (July 18, 2017); Human Rights Council, *Report of the UN High Commissioner for Human Rights on Improving Accountability and Access to Remedy for Victims of Business-Related Human Rights Abuse*, U.N. Doc. A/HRC/32/19, ¶¶ 4–6 (May 10, 2016).

174. *See, e.g.*, Ramsden, *Targeted Killings*, *supra* note 166, at 392–94; J.-G. Castel & Matthew E. Castel, *The Road to Artificial Superintelligence: Has International Law a Role to Play?*, 14(1) CANADIAN J. L. & TECH. 1, at 5 (2016) (discussing the threat to life by state-sponsored AI development and use).

law, to take precautions to address AI as an existential threat. At the same time, there is some movement toward a recognition of private actors being subjected to human rights obligations. While this has yet to emerge as a matter of international law, it is worthwhile setting out here this future possibility and its implications for AI design and use.

In this regard, the current vertical nature of human rights obligations (i.e., applicable to states and not private actors) seems increasingly out of place with the rise of "big tech" multinational corporations and their intention to develop AI technology.[175] As a result of this and other harmful effects produced by corporations, there is increasing awareness that private enterprises are capable of infringing human rights, with the emergence of a set of principles to ensure respect for human rights, as contained in the U.N. Guiding Principles on Business and Human Rights.[176] This "responsibility to respect" human rights requires that business enterprises avoid causing or contributing to adverse human rights impacts through their own activities, address such impacts when they occur, and seek to prevent or mitigate adverse human rights impacts directly linked to their operations.[177] However, the problem is that this does not impose a legal obligation, leaving it up to corporations to decide the extent to which they wish to incorporate human rights compliance in their practices, at least as a matter of law.[178] In failing to impose obligations on corporations directly, international human rights law thus does not fully capture the modern realities of power given the outsized role private corporations play in the development of AI. It might be that, over time, international human rights law develops to impose legal obligations directly on corporations; however, that point has not been reached yet.[179]

---

175. Indeed, the big tech monopoly over AI has been noted in the scholarly literature. *See* Pieter Verdegem, *Dismantling AI capitalism: the commons as an alternative to the power concentration of Big Tech*, 39 AI & Soc. 727 (2021) (discussing the "political economy of AI capitalism"); *see also* Christophe Samuel Hutchinson, *Potential abuses of dominance by big tech through their use of Big Data and AI*, 10(3) J. Antitrust Enf't 443 (2022) (discussing tech companies' potential to use AI and big tech to engage in anti-competitive behavior).

176. U.N. Special Representative of the Secretary-General, *Guiding Principles on Business and Human Rights: Implementing the United Nations "Protect, Respect and Remedy" Framework*, at 6, U.N. Doc. A/HRC/17/31 (Mar. 21, 2011). See also *Explanatory Report to the Council of Europe Framework Convention on Artificial Intelligence, supra* note 15, item 68 (noting that "all actors responsible for the activities within the lifecycle of artificial intelligence systems, irrespective of whether they are public or private organisations, must be subject to each Party's existing framework of rules, legal norms and other appropriate mechanisms so as to enable effective attribution of responsibility applied to the context of artificial intelligence systems").

177. *Id.* at 15.

178. Charamba, *supra* note 173, at 113.

179. *See e.g.*, Michael Ramsden, *Collective Legalization as a Strategic Function of the UN General Assembly in Responding to Human Rights Violations*, 40(1) Wis. Int'l L. J. 51, 84

As such, there are limitations to using international human rights law as a constraint on private entities engaged in the development of AI *directly*. Insofar as AI programs have been designed under the purview of state actors, there is a direct obligation upon such actors to ensure that any such design and use of AI is proportionate to human rights, including the right to life. Yet, despite these limitations with negative obligation, the following section will argue that the state is under a positive obligation to apply the precautionary approach irrespective of whether public or private enterprises control the design of AI systems.

### C.    *States' Positive Obligation to Prevent AI as a Threat to Life*

While there are limitations in imposing direct obligations on private enterprises to observe human rights, it is still incumbent on the state to adopt measures that ensure such enterprises observe rights. This leads to the positive obligation on the part of states to prevent deprivations of life, and, as this section develops, to address potential threats to life. As will be further developed below, the nature of this obligation varies but generally requires that the state "take adequate measures of protection, including continuous supervision, in order to *prevent. . .arbitrary deprivation of life*."[180] Before addressing the core issue here—concerning the degree to which evidence of an existential threat is required to trigger the positive obligation to address that threat—there are two preliminary points pertinent to the AI threat that must first be discussed.

#### 1. Non-Human Threats and Transboundary Considerations

First, it is not necessary for the threat to life to be attributed to a human or entity controlled by humans to trigger a positive obligation on states to protect life. For example, the European Court of Human Rights has held that a natural disaster triggered a state's obligation to prevent the threats to life arising from the disaster.[181] The Human Rights Committee's General Comment No. 36 similarly lists life-threatening diseases and environmental degradation as other threats to life that require state action under the right to life.[182] Given that the

---

(2023) (discussing the UN's use of collective legalization to address violations of international law).

180. General Comment 36, *supra* note 157, ¶ 21 (emphasis added).

181. Budayeva and others v. Russia, App. Nos. 15339/02, 21166/02, 20058/02, 11673/02 and 15343/02, ¶ 132 (Mar. 20, 2008).

182. The Human Rights Committee has set out, non-exhaustively, different types of threats to life that trigger the positive obligation on the state to act: See General Comment 36, *supra* note 157, ¶ 26 (including "high levels of criminal and gun violence, pervasive traffic and industrial accidents, degradation of the environment, deprivation of indigenous peoples' land, territories and resources, the prevalence of life-threatening diseases, such as AIDS, tuberculosis and malaria, extensive substance abuse, widespread hunger and malnutrition and extreme poverty and homelessness" (footnotes excluded)).

overarching focus is on threats to life rather than on the form this threat takes, the positive obligation would naturally extend to the threat posed by AI, be that at the design phase or even in a scenario where humans had lost control over a misaligned rogue AI system.

Second, it is also apparent that the failure to observe a positive obligation, in the context of transboundary harms, would make a state responsible for a threat to life that manifests itself in the territory of another state: here a state has an obligation to address the transboundary threat arising from its territory or subject to its jurisdiction.[183] In the context of transboundary environmental damage arising from climate change, the Inter-American Court of Human Rights thus observed that states "can be held responsible for significant damage caused to persons located outside their territory as a result of activities originating in their territory or under their authority or effective control."[184] Clearly, this extraterritorial principle applies to any threat to life originating from the territory of a state and would thus include the threat posed by the development of AI technology within a state.

### 2. Addressing Uncertain Threats

However, the more salient issue in the context of the threat to life posed by AI is the extent to which there needs to be evidence of this threat as a matter of international human rights law to trigger the positive obligation for the state to act. A restrictive approach in this respect would be to confine the positive obligation to one of preventing threats that are evidentially certain, or which are otherwise imminent. On this basis, in the particular context of law enforcement in averting threats to targeted individuals, the European Court of Human Rights has noted that it must be established that the party violating its obligation "knew or ought to have known at the time of the existence of a real and immediate risk to the life" of such individuals.[185] On this basis, the complainant must furnish evidence to establish a direct

---

183. Osman v. United Kingdom, App. No. 23452/94, ¶ 115 (Oct. 28, 1998) https://hudoc.echr.coe.int/eng#{%22itemid%22:[%22001-58257%22]}; General Comment 36, *supra* note 157, ¶ 22; Yassin v. Canada, U.N. Doc. CCPR/C/120/D/2285/2013, ¶ 6.5 (Jul. 26, 2017); Human Rights Comm., Concluding Observations on the Sixth Periodic Report on Canada, ¶ 6, U.N. Doc. CCPR/C/CAN/CO/6 (Aug. 13, 2015).

184. Advisory Opinion on Human Rights and the Environment, Advisory Opinion OC-23/17, Inter-Am. Ct. H.R. (ser. A), ¶¶ 101–03 (Nov. 15, 2017); *see also* Human Rights Comm., 118th session, 3323rd meeting (Oct. 26, 2016), http://webtv.un.org/en/asset/k1t/k1t5oka5q9 (no transcript available); *see also* Human Rights. Comm., General Comment No. 31, The Nature of the General Legal Obligation Imposed on States Parties to the Covenant, U.N. Doc. CCPR/C/21/Rev.1/Add.13, ¶ 8 (May 26, 2004).

185. Osman v. United Kingdom, App. No. 23452/94, ¶ 116 (Oct. 28, 1998) https://hudoc.echr.coe.int/eng#{%22itemid%22:[%22001-58257%22]}.

connection between the threatening activity and the alleged rights violation.[186]

Yet this restrictive approach—premising the exercise of a positive obligation on the existence of a real and immediate risk to life—would reduce the utility of international human rights law in addressing grave risks where there remains uncertainty as to the extent to which that risk will materialize. Given this, a broader approach is to construe the positive obligation to protect life in relation to broader threats to society at large and those from which there is a lack of evidentiary consensus over the nature of the threat—in short, the precautionary principle outlined in Part III. Indeed, in moving beyond the more discrete mandate of protecting individuals from immediate threats to them, it is apparent from jurisprudential developments that the positive obligation under the right to life has evolved to cover grave risks to life that remain uncertain but nonetheless require positive state action.

The Human Rights Committee's General Comment No. 36, adopted in 2019, is a useful starting point in this regard. There, the positive obligation was framed as requiring the state to exercise due diligence in response to "reasonably foreseeable threats and life-threatening situations that can result in loss of life" or otherwise a "real risk of irreparable harm."[187] As this formulation implies, it is not necessary for this threat to be imminent. In *Daniel Billy v. Australia*, the Human Rights Committee thus considered potential threats to life arising from sea level rise 10-15 years in the future.[188] Indeed, the threat need not materialize at all: state parties may be in violation of the right to life "even if such threats and situations do not result in loss of life."[189] Perhaps the most significant statement of all in General Comment No. 36 was the explicit endorsement of the "precautionary approach," at least in the context of "[e]nvironmental degradation, climate change and unsustainable development," all of which "constitute some of the most pressing and serious threats to the ability of present and future generations to enjoy the right to life."[190]

---

186. Balmer-Schafroth and others v. Switzerland, App. No. 67/1996/686/876 (Aug. 26, 1997) https://hrcr.org/safrica/environmental/balmer_switzerland.html (noting that the complainants were not placed in "specific, grave, and imminent danger" with the renewal of a power station licence in their vicinity).

187. General Comment 36, *supra* note 157, at 7. This principle has also been applied in decisions of the Human Rights Committee. *See, e.g.*, Toussaint v. Canada, U.N. Doc. CCPR/C/123/D/2348/2014, Human Rights Comm., Decision, ¶ 11.3 (July 24, 2018); Portillo Cáceres v. Paraguay, U.N. Doc. CCPR/C/126/D/2751/2016, Human Rights Comm., Decision,¶ 7.5 (July 25, 2019); NS v. Russian Federation, U.N. Doc. CCPR/C/113/D/2192/2012, Decision, ¶ 10.4 (June 1, 2015); Teitiota v. New Zealand, CCPR/C/127/D/2728/2016, Human Rights Comm., Decision, ¶ 9.3 (Sept. 23, 2020); UN Human Rights Council, Report of The Special Representative of The Secretary-General on The Issue of Human Rights and Transnational Corporations and Other Business Enterprises, U.N. Doc. A/HRC/17/31, principles, at 1–10 (Mar. 21, 2011).

188. Daniel Billy v. Australia, U.N. Doc. CCPR/C/135/D/3624/2019, Human Rights Comm., Decision, ¶ 8.7 (Sept. 23, 2022).

189. General Comment 36, *supra* note 157, ¶ 7.

The precautionary principle has similarly been recognized in the construction of the right to life in regional human rights systems. The Inter-American Court of Human Rights has thus opined that "[s]tates must act in keeping with the precautionary principle to protect the rights to life in the event of possible serious and irreversible damage to the environment, even in the absence of scientific certainty."[191] A line of authorities in the European Court of Human Rights apply a precautionary approach, even if the principle is not always explicitly referenced.[192] In particular, the Court has held that a positive obligation to protect life indisputably exists in relation to "dangerous activities" that only pose a potential threat to life that had yet to materialize.[193] These included, for example, the operation of a waste-collection site or a man-made reservoir close to human populations.[194] The positive obligation has extended to the public health sphere, requiring the state to impose requirements on hospitals to take appropriate measures to ensure the protection of human life, even in the absence of any specific threat or violation.[195]

These iterations reflect the need to avoid a restrictive interpretation of the right to life focused on evidence and proof of violations or imminent threats to life; to the contrary, the right needs to be interpreted and applied in a manner as to make its safeguards practical and effective.[196] Thus, as Yuval Shany, the lead author of General Comment No. 36 noted, "to effectively protect the right to life, it is not possible to intervene at the very last minute and states need to deal with the *causes* of the violation."[197] To not deal with the causes, according to Shany, would be like "closing the stable door after the horse has bolted."[198] In turn, the interpretive principle of effectiveness supports the general proposition (beyond the defined applications above) that precaution is a necessary element of the state's positive obligation in relation to large-scale and grave but uncertain threats to human life.[199]

---

190. *Id.* ¶ 62.

191. Advisory Opinion on Human Rights and the Environment, Advisory Opinion OC-23/17, Inter-Am. Ct. H.R. (ser. A), ¶¶ 101–103 (Nov. 15, 2017).

192. Tătar c. Roumanie, App. No. 67021/01, ¶ 120 (July 6, 2009).

193. Kolyadenko and others v. Russia, App. Nos. 17423/05, 20534/05, 20678/05, 23263/05, 24283/05 and 35673/05, ¶ 164 (Feb. 28, 2012) https://hudoc.echr.coe.int/eng#{%22itemid%22:[%22001-109283%22]}; Öneryıldız v. Turkey, App. No. 48939/99, ¶ 71 (Nov. 30, 2004); Tătar v. Romania, App No. 67021/01 (Jan. 27, 2009), https://hudoc.echr.coe.int/eng#{%22itemid%22:[%22001-90909%22]}; s*ee also* Case C-180/96, United Kingdom of Great Britain and Northern Ireland v. Comm'n, 1998 E.C.R I-02265, ¶ 99 ("Where there is uncertainty as to the existence or extent of risks to human health, the institutions may take protective measures without having to wait until the reality and seriousness of those risks become fully apparent.").

194. Kolyadenko and others v. Russia, App. Nos. 17423/05, 20534/05, 20678/05, 23263/05, 24283/05 and 35673/05, ¶ 164 (Feb. 28, 2012) https://hudoc.echr.coe.int/eng#{%22itemid%22:[%22001-109283%22]}; Öneryıldız v. Turkey, App. no. 48939/99, ¶ 71 (Nov. 30, 2004).

195. Calvelli and Ciglio v. Italy, App. No. 32967/96, ¶ 49 (Jan. 27, 2002), https://hudoc.echr.coe.int/eng#{%22itemid%22:[%22001-60329%22]}; Ciechońska v. Poland, App. No. 19776/04, ¶ 69 (June 14, 2011), https://hudoc.echr.coe.int/eng#{%22itemid%22:[%22001-105102%22]}.

196. *See* General Comment 36, *supra* note 157, ¶ 3 (arguing that the right to life "should not be interpreted narrowly"); Öneryıldız v. Turkey, App. no. 48939/99, ¶ 69 (Nov. 30, 2004).

### 3. The Positive Obligation of States to Regulate AI Activities

All of this supports the interpretive claim that AI, as a technology posing potential existential risk, with no evidential certainty or scientific consensus as to the nature of this threat, nonetheless compels states to act. But what is the content of this positive obligation? International decision-makers have defined it to vary depending on the activity in question, taking into account the state's margin of appreciation to choose the appropriate means for discharging this obligation.[200] At the same time, the core of this obligation is that states have an ongoing duty to assess the potential risks inherent in the activity and to "place a legislative and administrative framework designed to provide effective deterrence against threats to the right to life."[201] Such legislative and administrative framework would include regulating "the licensing, setting up, operation, security and supervision of the activity and must make it compulsory for all those concerned to take practical measures to ensure the effective protection of citizens whose lives might be endangered by the inherent risks."[202]

---

197. Human Rights Comm., 3323rd meeting, *supra* note 184 (emphasis added); *see also* discussion in Ginevra Le Moli, *The Human Rights Committee, Environmental Protection and the Right to Life*, 69 INT'L & COMP. L. Q. 735, 742 (2020).

198. Human Rights Comm., 3323rd meeting, *supra* note 184.

199. As to the principle of effectiveness, *see* Georgios A. Serghides, *The Principle of Effectiveness and its Overarching Role in the Interpretation and Application of the European Convention on Human Rights, in Particular Its Relationship to the Other Convention Principles,* 30 HAGUE Y.B. INT'L L. 1 (2017) (discussing the "immense importance" the principle of effectiveness has in the European Convention on Human Rights).

200. Case of Budayeva and others v. Russia, App. Nos. 15339/02, 21166/02, 20058/02, 11673/02 & 15343/02, ¶ 134 (Mar. 20, 2008), https://hudoc.echr.coe.int/eng#{%22itemid%22:[%22001-85436%22]}.

201. *Id.* ¶ 128; Kolyadenko and Others v. Russia, App. Nos. 17423/05, 20534/05, 20678/05, 23263/05, 24283/05 & 35673/05, ¶¶ 157, 166 (Feb. 28, 2012), https://hudoc.echr.coe.int/eng#{%22itemid%22:[%22001-109283%22]}; Calvelli and Ciglio v. Italy, App. No. 32967/96, ¶ 49 (Jan. 17, 2002), https://hudoc.echr.coe.int/eng#{%22itemid%22:[%22001-60329%22]}.

202. Case of Budayeva and others v. Russia, App. Nos. 15339/02, 21166/02, 20058/02, 11673/02 & 15343/02, ¶ 132 (Mar. 20, 2008), https://hudoc.echr.coe.int/eng#{%22itemid%22:[%22001-85436%22]}.

It extends to ensuring an effective remedy is provided where the right to life is violated, holding those at fault accountable and providing appropriate redress to the victim(s).[203] Given that at the conceptualization stage the full capabilities of AI cannot be predicted, some scholars have, in turn, considered the scope for reduced human responsibility in the event AI violates human rights and other legal norms.[204] Yet, the fact remains that actors will still be designing AI, knowing that they cannot predict the effect it will have.[205] That the designers cannot anticipate the level of risk due to the complexity of the AI system would not justify a reduction in responsibility.[206] If the actor, notwithstanding the grave risk, still proceeded with AI design, such a decision should incur heightened responsibility.[207] While this point may appear moot if a super intelligent AI system went on to end humanity, attacks that fall short of this would trigger a positive obligation on the state to investigate and prosecute those humans responsible for the AI system in question.[208] The positive obligation to criminal legal rules governing violations of human rights and the right to life thus provides a possible constraint upon developers and thus an additional tool in the effort to ensure precaution in AI design.

## V. Conclusion

AI experts generally split into two camps: those who are optimistic about the contributions that AI will bring and are reasonably sanguine regarding its risks, and those who are extremely alarmed at the potential threat AI poses. However, the AI community is unified in its belief that paradigm-shifting advancements in AI technology are imminent. Given the potential for significant harm and the absence of scientific certainty surrounding AI risks, this article argued that the precautionary principle obligates states to regulate this powerful new technology, as failing to or even delaying doing so could result in potentially irreversible catastrophic consequences.

---

203. Centre for Legal Resources on behalf of Valentin Câmpeanu v. Romania, App. no. 47848/08, ¶ 132 (July 17, 2014).

204. McGregor et al., *supra* note 18, at 341; Brent Daniel Mittelstadt Patrick Allo, Mariarosaria Taddeo, Sandra Wachter & Luciano Floridi, *The Ethics of Algorithms: Mapping the Debate* Big Data & Soc. 1, 11–12 (2016); Andreas Matthias, *The Responsibility Gap: Ascribing Responsibility for the Actions of Learning Automata*, 6 Ethics & Info. Tech. 175, 177 (2004).

205. McGregor et al., *supra* note 18, at 341.

206. *Id.*

207. *Id.* at 341; Guido Acquaviva, *Crimes without Humanity?: Artificial Intelligence, Meaningful Human Control, and International Criminal Law*, 21 J. Int'l Crim. Just. 981 (2023).

208. *See further* Reinmar Nindler, *The United Nation's Capability to Manage Existential Risks with a Focus on Artificial Intelligence*, 21 Int'l Comty. L. Rev. 5, 22 (2019); Burri, *supra* note 169, at 108.

It was argued that the precaution is a general principle of international law, and that, as such, international human rights law imposes a positive obligation on states under the right to life to proactively take regulatory action to safeguard human life in the face of the possible existential threat posed by machine intelligence. As it was beyond the scope of the discussion, the article, however, did not delve into the specific forms such regulation could take. The authors strongly invite examination of this kind. The goal of this article was to take the first crucial step and establish a legal basis for an international law response to the existential risk associated with AI technology with the hope that doing so will help the international community marshal an effective legal response to this evolving threat.